\documentclass[lettersize,journal]{IEEEtran}
\usepackage{amsmath,amsfonts}
\usepackage{algorithmic}
\usepackage{array}
\usepackage[caption=false,font=scriptsize,labelfont=rm,textfont=rm]{subfig}
\usepackage{textcomp}
\usepackage{stfloats}
\usepackage{threeparttable}
\usepackage{url}
\usepackage{verbatim}
\usepackage{graphicx}
\def\BibTeX{{\rm B\kern-.05em{\sc i\kern-.025em b}\kern-.08em
		T\kern-.1667em\lower.7ex\hbox{E}\kern-.125emX}}
\usepackage{balance}

\usepackage{multirow}
\usepackage{cite}
\usepackage{cases}
\usepackage{amssymb}
\usepackage[ruled]{algorithm2e}
\usepackage{bm}
\usepackage{color,xcolor}

\usepackage{amsthm}

\newtheorem{mypro}{Proposition}

\allowdisplaybreaks[3]

\begin{document}
	
	\title{Multi-Waveguide Pinching Antennas for ISAC}
    
	\author{Weihao Mao,~\IEEEmembership{Student Member,~IEEE}, Yang Lu,~\IEEEmembership{Member,~IEEE}, Yanqing Xu,~\IEEEmembership{Member,~IEEE}, \\
  Bo Ai,~\IEEEmembership{Fellow,~IEEE}, Octavia A. Dobre,~\IEEEmembership{Fellow,~IEEE}, and Dusit Niyato,~\IEEEmembership{Fellow,~IEEE}
		\thanks{Weihao Mao and Yang Lu are with the State Key Laboratory of Advanced Rail Autonomous Operation, and also with the School of Computer Science and Technology, Beijing Jiaotong University, Beijing 100044, China (e-mail: weihaomao@bjtu.edu.cn, yanglu@bjtu.edu.cn).}
		\thanks{Yanqing Xu  is with the School of Science and Engineering, The Chinese University of Hong Kong, Shenzhen, China. (e-mail: xuyanqing@cuhk.edu.cn)}
        \thanks{Bo Ai is with the School of Electronics and Information Engineering, Beijing Jiaotong University, Beijing 100044, China (email:  boai@bjtu.edu.cn).}
        \thanks{Octavia A. Dobre is with the Faculty of Engineering and Applied Science, Memorial University, St. John’s, NL A1C 5S7, Canada (E-mail: odobre@mun.ca).}
        \thanks{Dusit Niyato is with the College of Computing and Data Science, Nanyang Technological University, Singapore 639798 (e-mail: dniyato@ntu.edu.sg).}
        \thanks{This work has been submitted to the IEEE for possible publication. Copyright may be transferred without notice, after which this version may no longer be accessible.}
        }
	
	\maketitle
	\thispagestyle{empty}
	
	\begin{abstract}
        Recently, a novel flexible-antenna technology, called  pinching antennas, has attracted growing academic interest. By inserting discrete  dielectric materials, pinching antennas can be activated at arbitrary points along waveguides, allowing for flexible customization of large-scale path loss. This paper investigates a multi-waveguide pinching-antenna  integrated sensing and communications (ISAC) system, where  transmit pinching antennas (TPAs) and receive pinching antennas (RPAs) coordinate to simultaneously detect one potential target and serve one downlink user. We formulate a communication rate maximization problem subject to radar signal-to-noise ratio (SNR) requirement, transmit power budget, and the allowable movement region of the TPAs, by jointly optimizing TPA locations and transmit beamforming design. To address the non-convexity of the problem, we propose a novel fine-tuning approximation method to reformulate it into a tractable form, followed by a successive convex approximation (SCA)-based algorithm to obtain the solution efficiently. Extensive simulations validate both the system design and the proposed algorithm. Results show that the proposed method achieves near-optimal performance compared with the computational-intensive exhaustive search-based benchmark, and pinching-antenna ISAC systems exhibit a distinct communication-sensing trade-off compared with conventional systems.
        \end{abstract}
	
	
	\IEEEpeerreviewmaketitle

	\setlength{\parindent}{1em}
	
	\section{Introduction}
	
	The future sixth-generation (6G) mobile communication systems are expected to realize both ubiquitous connectivity and hyper-reliable low-latency communication \cite{6g}. However, providing high quality coverage remains challenging when users are obstructed or located at cell edges. To address this issue, flexible-antenna technologies have garnered  increasing attention, e.g., reconfigurable intelligent surface \cite{ris} (RISs, also known as intelligent reflecting surfaces (IRS) \cite{irs}),  fluid-antenna system \cite{fas}, and movable-antenna system \cite{mas}. Compared with fixed-location antenna systems, these flexible-antenna systems enable programmable and controllable wireless channels at the small-scale level to improve the channel conditions, e.g., fine-tuning the location of antennas or phase shifts of reflecting elements. Nevertheless, most existing flexible-antenna technologies do not support adjustments to  the large-scale path loss, which dominates the channel conditions. Recently, a novel flexible-antenna paradigm, termed as pinching antenna, has been proposed \cite{pinch}. Specifically, the pinching antenna systems employ multiple dielectric waveguides as transmission medium, and allow pinching antennas to be activated at arbitrary points along the waveguides via inserting  separate dielectric materials \cite{pinch2}. This enables a quick and flexible deployment of pinching antennas at the large-scale level. For instance, they can be deployed near users to reduce the propagation loss or establish line-of-sight (LoS) links \cite{pinch_ad1}. Leveraging the flexibility of pinching antennas to customize channel characteristics, some recent works have explored the potential of  pinching antenna systems to enhance mobile communications \cite{pinch_rw1, pinch_rw2, pinch_rw3, pinch_rw4}. In \cite{pinch_rw1}, the authors derived closed-form expressions for  communication outage probability and average rate of pinching antenna systems. Besides, \cite{pinch_rw2} derived a closed-form upper bound on array gain for pinching antenna systems with fixed inter-antenna spacing. Moreover, \cite{pinch_rw3} maximized the minimum transmission rate for uplink pinching antenna systems, while \cite{pinch_rw4} optimized the transmission rate for a single-user downlink pinching antenna system. The superiority of pinching antenna systems over conventional systems was demonstrated in aforementioned works in terms of reliability, array gain, and communication rates.

    On the other hand, the integrated sensing and communications (ISAC) technology has emerged  as a pivotal candidate for future 6G networks to accommodate the rapidly evolving application scenarios, such as smart cities and intelligent transportation \cite{ISAC_bc1}. Unlike conventional systems with separate base stations (BSs) for communication and standalone radars for sensing, ISAC enables the integration and joint optimization of dual functionalities within an ISAC BS, leveraging shared spectrum and hardware resources \cite{ISAC_bc2}. Beyond improving the spectral efficiency and hardware efficiency \cite{ISAC_bc3}, ISAC achieves  synergistic performance gains in both domains  through tightly coupled bidirectional interactions \cite{ISAC_bc4}. However, ISAC systems face a critical challenge, i.e., the sensing performance is highly dependent on LoS propagation condition, which is often disrupted by obstructions in practical scenarios \cite{ISAC_bc5}. To tackle this challenge, some works have investigated flexible-antenna ISAC systems, employing the benefit of establishing LoS links \cite{ISAC_rw1, ISAC_rw2, ISAC_rw3, ISAC_rw4}. In \cite{ISAC_rw1}, the authors investigated an active RIS-assisted ISAC system where the active RIS created a cascaded LoS link for sensing, and the radar signal-to-interference-plus-noise-ratio (SINR) was maximized subject to communication SINR requirements of users. Besides, \cite{ISAC_rw2} maximized the minimum of communication SINR and radar signal-to-clutter-and-noise-ratio (SCNR) in a simultaneously transmitting and reflecting (STAR)-RIS and non-orthogonal multiple access (NOMA) assisted ISAC system, demonstrating that STAR-RIS was essential for sensing in the non-LoS scenarios. Moreover, \cite{ISAC_rw3} maximized the downlink sum rate under the constraint of radar SINR requirements for fluid-antenna ISAC systems by optimizing the locations of antenna and beamformers. The authors of \cite{ISAC_rw4} maximized the weighted summation of communication rate and sensing mutual information (MI) via  flexible beamforming optimization in movable-antenna ISAC systems.

    While conventional flexible-antenna ISAC systems can enable LoS sensing, two critical  challenges persist. First, RISs excel at constructing LoS links but introduce multiplicative fading. Second, fluid-antenna and movable-antenna systems may fail to ensure LoS propagation in highly cluttered environments with dense obstacles. To address these limitations and expand  application scenarios, pinching-antenna ISAC systems have emerged as a promising solution, offering simultaneous  customization of both large-scale and small-scale channel  characteristics \cite{pa_isac1, pa_isac2, pa_isac3, pa_isac4}. In \cite{pa_isac1}, the author derived the closed-form Cramér-Rao lower bound (CRLB) for pinching-antenna  ISAC systems and demonstrated that pinching antennas can significantly reduce the CRLB compared with conventional systems. Besides, \cite{pa_isac2} proposed a maximum entropy-based reinforcement learning approach to maximize the sum communication rate while satisfying the radar signal-to-noise ratio (SNR) requirements in pinching-antenna ISAC systems. Meanwhile, \cite{pa_isac3} investigated a pinching-antenna ISAC system with a transmit waveguide and a receive waveguide, and maximized the illumination power for sensing subject to communication requirements. Additionally, \cite{pa_isac4} characterized the performance region for pinching-antenna  ISAC systems, where the communication rate and the sensing rate were respectively employed as metrics. Notably, existing works \cite{pa_isac2, pa_isac3, pa_isac4} deployed all transmit pinching antennas (TPAs) on one waveguide, whereas a distributed architecture with TPAs across multiple waveguides can reduce average path loss over a larger coverage area \cite{cf}.   Besides, the optimization of TPA locations is challenging because they affect both path loss and phase shift of signals received at users, their optimization presents a significant challenge. Consequently, \cite{pa_isac3} and \cite{pa_isac4} employ exhaustive search algorithms for TPA placement, which are computationally prohibitive for multi-waveguide pinching-antenna ISAC systems, as the solution space grows exponentially with the number of waveguides.


    \emph{So far, efficient resource allocation for multi-waveguide pinching-antenna ISAC systems has remained an unaddressed research challenge.} To fill this gap, this paper proposes a computationally efficient algorithm for joint optimization of  TPA locations and transmit beamforming in multi-waveguide pinching-antenna ISAC systems. It is noted that compared with single-waveguide pinching-antenna ISAC systems, developing  resource allocation schemes for multi-waveguide pinching-antenna ISAC systems poses distinct challenges due to the expanded optimization space of TPA locations and the complicated relationship between TPA locations and channel characteristics. The main contributions of this paper are outlined as follows:
    \begin{enumerate}
        \item
        We consider  a multi-waveguide pinching-antenna ISAC system, where multiple TPAs are coordinated to simultaneously provide communication services to a user and illuminate a potential target, while multiple receive pinching antennas (RPAs) acquire the reflected signals to perform target detection. We independently deploy each TPA and RPA on a dedicated waveguide. We  prove that the radar SNR is positively correlated with the detection probability and convert the complicated sensing probability requirement into a computationally tractable form.

        \item
        We derive the optimal  RPA locations through theoretical analysis and the optimal closed-form expression for  receive beamforming vector using the Cauchy-Schwarz inequality. Building on these insights, we formulate a communication rate maximization problem subject to radar SNR requirement, the transmit power budget, and the allowable movement region of the TPAs, by jointly optimizing TPA locations and the transmit beamforming design.

        \item
         To address the non-convexity of the problem, we first reformulate it based on the derived optimal solution structure, and then present a novel fine-tuning method to approximate it with negligible performance loss. By introducing auxiliary variables,  the approximated problem is further transformed into a computationally tractable form, which is then efficiently solved using a proposed successive convex approximation (SCA)-based algorithm. 

        \item 
        Numerical results validate the effectiveness of the proposed system and algorithm. It is shown that the proposed SCA-based algorithm achieves performance nearly identical to the computational-intensive exhaustive search benchmark. Furthermore, the proposed pinching-antenna ISAC design outperforms both the conventional fixed-antenna scheme and three representative pinching-antenna benchmarks in terms of communication rate under radar SNR constraints. In addition, the communication rate exhibits non-smooth variations as the radar SNR requirement increases, clearly illustrating the fundamental communication-sensing trade-off in  pinching-antenna ISAC systems. 
    \end{enumerate}

    The remainder of this paper is structured as follows. Section II introduces the multi-waveguide pinching-antenna ISAC system model and formulates a communication rate maximization problem under radar SNR constraint. Section III presents a fine-tuning approximation for the optimization problem, which is subsequently solved using an SCA-based algorithm. Section IV provides simulation results and discussions to evaluate the performance of the proposed algorithm and the effectiveness of the pinching-antenna ISAC system. Section V concludes this paper.

    {\it Notations:} In this paper, $x$, ${\bf x}$, ${\bf X}$ and ${\cal X}$ are respectively denoted by scalar, vector, matrix and set. ${\rm Re}\{ \cdot  \}$ denotes the real part of a complex number, vector or matrix. $|| \cdot ||$ denotes the two-norm for a complex vector and $| \cdot |$ denotes the magnitude for a complex scalar. $(\cdot)^T$, $(\cdot)^*$ and $(\cdot)^H$ represent the transpose, conjugate and conjugate transpose, respectively. ${\mathbb C}^M$ and  ${\mathbb C}^{M \times N}$ denote the set of $M \times 1$ complex-valued vectors and $M \times N$ complex-valued matrices, respectively. ${\bf I}_N$ and ${\bf 0}_N$ are the $N$-dimension identity matrix and zero vector, respectively. ${\bf a} \sim \mathcal{CN}({\bm \mu}, {\bm \Sigma})$ denotes that ${\bf a}$ is a complex-valued circularly symmetric Gaussian random variable with mean  ${\bm \mu}$ and covariance matrix ${\bm \Sigma}$.

	\section{System Model}
	
	\begin{figure}
		\begin{center}
			\centerline{\includegraphics[ width=.49\textwidth]{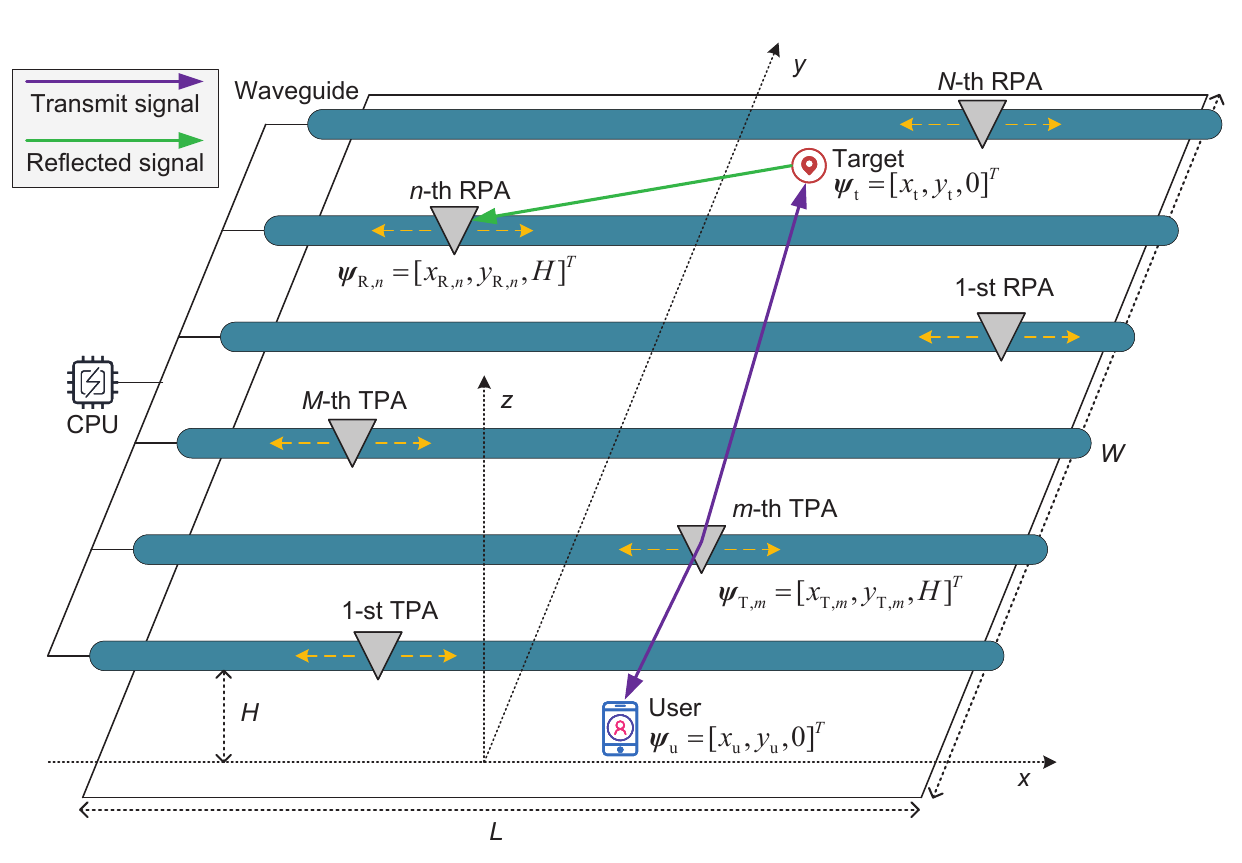}}
			\caption{Illustration of a multi-waveguide  pinching-antenna ISAC system.}
			\label{sys}
		\end{center}
		\vspace{-0.8cm}
	\end{figure}
	Consider a multi-waveguide pinching-antenna  ISAC system as shown in Fig. \ref{sys}, where a central processing unit (CPU) coordinates $M$ TPAs and $N$ RPAs to jointly serve one single-antenna downlink user and sense one point-like target.
	The user and the target are located within a rectangular region with dimensions of length $L$ and width $W$, while each TPA/RPA is located on one separate transmit/receive waveguide, i.e., the $m$-th TPA ($m \in \mathcal{M} \triangleq \{1,2,\ldots, M\}$) is located on the $m$-th transmit waveguide and the $n$-th RPA ($n \in \mathcal{N} \triangleq \{1,2,\ldots, N\}$) is located on the $n$-th receive waveguide. Particularly, each TPA/RPA is movable on its waveguide with a scale much larger than a wavelength such that it can be deployed next to the user or the target.
	
	A three-dimensioned Cartesian coordinate system is built to describe the considered system. The coordinates of the $m$-th TPA and the $n$-th RPA are respectively denoted by ${\bm \psi}_{{\rm T}, m} = [x_{{\rm T}, m}, y_{{\rm T}, m}, H]^T$ and ${\bm \psi}_{{\rm R}, n} = [x_{{\rm R}, n}, y_{{\rm R}, n}, H]^T$, where $y_{{\rm T}, m}$,  $y_{{\rm R}, n}$, and $H$ are pre-determined according to the locations of waveguides, while $x_{{\rm T}, m}$ and $x_{{\rm R}, n}$ are optimization variables. The user and the target are fixed on the ground, and their coordinates are respectively denoted by ${\bm \psi}_{\rm u} = [x_{\rm u}, y_{\rm u}, 0]^T$ and  ${\bm \psi}_{\rm t} = [x_{\rm t}, y_{\rm t}, 0]^T$.

	\subsection{Channel Model}
	Following the classical spherical wave channel model\cite{channel_model}, the channel from the $m$-th TPA to the user is given by
	\begin{flalign}
		h_{{\rm u}, m} \left( x_{{\rm T}, m} \right) = \frac{\sqrt{\eta} e^{ - 2 \pi j \left( \frac{\left|  \left| {\bm \psi}_{\rm u} -  {\bm \psi}_{{\rm T}, m}  \right| \right|}{\lambda} + \frac{\left|  \left| {\bm \psi}_{{\rm T},m}^{0} -  {\bm \psi}_{{\rm T}, m}  \right| \right|}{\lambda_g}    \right)  }  }{ \left|  \left| {\bm \psi}_{\rm u} -  {\bm \psi}_{{\rm T}, m}  \right| \right|   }, \label{h_u}
	\end{flalign}
	where $\eta = c^2/ (4 \pi f_c)^2$ with $c$ denoting the speed of light and $f_c$ denoting  the carrier frequency, $\lambda = c/f_c$ and $\lambda_g = c / (f_c n_{\rm eff}) $, with $ n_{\rm eff}$ denoting the effective refractive index of a dielectric waveguide \cite{neff}. Here, ${\bm \psi}_{{\rm T}, m}^0 = [-L/2, y_{{\rm T}, m}, H]^T$ is pre-defined, which denotes the location of the feed point of the $m$-th transmit waveguide. 
	
	Note that, in \eqref{h_u},  the phase shift $e^{ - 2 \pi j  \left|  \left| {\bm \psi}_{\rm u} -  {\bm \psi}_{{\rm T}, m}  \right| \right| / {\lambda}   }$ is caused by the free-space propagation of signals from the $m$-th TPA to the user, while the phase shift $e^{- 2 \pi j || {\bm \psi}_{{\rm T}, m}^0 - {\bm \psi}_{{\rm T}, m}  || / \lambda_g}$ is caused by the propagation of signals inside the $m$-th transmit waveguide.
	
	Similarly, the channels from the $m$-th TPA to the target and from the $n$-th RPA to the target are respectively given by
	\begin{flalign}
		& h_{{\rm t}, m} \left( x_{{\rm T}, m} \right) = \frac{\sqrt{\eta} e^{ - 2 \pi j \left( \frac{\left|  \left| {\bm \psi}_{\rm t} -  {\bm \psi}_{{\rm T}, m}  \right| \right|}{\lambda} + \frac{\left|  \left| {\bm \psi}_{{\rm T},m}^{0} -  {\bm \psi}_{{\rm T}, m}  \right| \right|}{\lambda_g}    \right)  }  }{ \left|  \left| {\bm \psi}_{\rm t} -  {\bm \psi}_{{\rm T}, m}  \right| \right|   }, \label{h_t}  \\
		& g_{{\rm t}, n} \left( x_{{\rm R}, n} \right) = \frac{\sqrt{\eta} e^{ - 2 \pi j \left( \frac{\left|  \left| {\bm \psi}_{\rm t} -  {\bm \psi}_{{\rm R}, n}  \right| \right|}{\lambda} + \frac{\left|  \left| {\bm \psi}_{{\rm R},n}^{0} -  {\bm \psi}_{{\rm R}, n}  \right| \right|}{\lambda_g}    \right)  }  }{ \left|  \left| {\bm \psi}_{\rm t} -  {\bm \psi}_{{\rm R}, n}  \right| \right|   }, \label{g_t} 
	\end{flalign}
	where  ${\bm \psi}_{{\rm R}, n}^0 = [-L/2, y_{{\rm R}, n}, H]^T$ is pre-defined and denotes the location of the feed point of the $n$-th receive waveguide \cite{pinch}.

	\subsection{Communication Model}
	The signal transmitted by the TPAs is given by ${\bf x} = {\bf w} s$, where $s \in \mathbb{C}$ with $\mathbb{E}\{ |s|^2 \} = 1$ denotes the symbol transmitted to the user and ${\bf w} \in \mathbb{C}^M$ denotes the corresponding beamforming vector. 
	
	Then, the received signal at the user is given by
	\begin{flalign}
		y_{\rm c} = {\bf h}_{\rm u}^H \left(\{ x_{{\rm T}, m}    \}   \right) {\bf w} s + n_{\rm u},
	\end{flalign}
	where ${\bf h}_{\rm u} ( \{ {x_{{\rm T}, m}}  \}) \triangleq [h_{{\rm u}, 1}(x_{{\rm T}, 1}), \ldots, h_{{\rm u}, M}(x_{{\rm T}, M})]^T$ and $n_{\rm u} \sim \mathcal{CN}(0, \sigma_{\rm u}^2) $ denotes the additive white Gaussian noise (AWGN) at the user. The communication SNR of the user is 
	\begin{flalign}
		{\Gamma}_{\rm u} \left(\{{\bf w}, x_{{\rm T}, m} \}  \right) = \frac{|{\bf h}_{\rm u}^H \left( \{ x_{{\rm T}, m}  \} \right) {\bf w} |^2}{\sigma_{\rm u}^2}.
	\end{flalign}

	\subsection{Sensing Model}
	Target detection focuses on describing the presence of the potential target, which is a typical task for ISAC. In fact, the target detection task is implemented based on the reflected signals received by RPAs, which is given by 
	\begin{flalign}
		{\bf y}_{\rm s} = \alpha {\bf g}_{\rm t} \left( \{ x_{{\rm R}, n}   \}  \right) {\bf h}_{\rm t}^H \left( \{ x_{{\rm T}, m}   \}  \right) {\bf w} s  + {\bf n}_{\rm s},
	\end{flalign}
	where $\alpha$ denotes the reflection coefficient of the target, ${\bf g}_{\rm t}(\{ x_{{\rm R}, n} \}) \triangleq [g_{{\rm t}, 1} ( x_{{\rm R}, 1}), \ldots,  g_{{\rm t}, N} ( x_{{\rm R}, N}) ]^T $, ${\bf h}_{\rm t}(\{ x_{{\rm T}, m} \}) \triangleq [ h_{{\rm t}, 1} ( x_{{\rm T}, 1} ), \ldots,  h_{{\rm t}, M} ( x_{{\rm T}, M})]^T$, and ${\bf n}_{\rm s} \in \mathcal{CN} ({\bf 0}_{N}, \sigma_{\rm s}^2 {\bf I}_{N} ) $ denotes the AWGN at the RPAs. 
	
	To enhance the detection probability, the RPAs strengthen ${\bf y}_s$ via the receive beamforming technology, i.e., 
	\begin{flalign}
		\widetilde{y}_{\rm s} &\triangleq {\bf v}^H {\bf y}_{\rm s} \nonumber\\
		&=\alpha {\bf v}^H {\bf g}_{\rm t} \left( \{ x_{{\rm R}, n}   \}  \right) {\bf h}_{\rm t}^H \left( \{ x_{{\rm T}, m}   \}  \right) {\bf w} s  + {\bf v}^H {\bf n}_{\rm s}, \label{receive_beam}
	\end{flalign}
	where ${\bf v} \in \mathbb{C}^N$ denotes the receive beamforming vector. 
	
	Then, the binary hypothesis testing problem for the target detection task can be formulated as \cite{radar_snr}:
	
	\begin{subnumcases}
		{\widetilde{y}_{\rm s} = }
		{\mathcal{H}_0}:  {\bf v}^H {\bf n}_{\rm s}, \\
		{\mathcal{H}_1}: \alpha {\bf v}^H {\bf g}_{\rm t} \left( \{ x_{{\rm R}, n}   \}  \right) {\bf h}_{\rm t}^H \left( \{ x_{{\rm T}, m}   \}  \right) {\bf w} s  + {\bf v}^H {\bf n}_{\rm s},
	\end{subnumcases}
	where the null hypothesis ${\mathcal{H}_0}$ represents the absence of the target, and the alternative hypothesis  ${\mathcal{H}_1}$ means that there exists a target. According to the linear transformation invariance of Gaussian distribution, we derive the conditional probability distributions $\widetilde{y}_{\rm s}|\mathcal{H}_{\rm 0} \sim \mathcal{CN}(0, \sigma_{\rm s}^2)$ and $\widetilde{y}_{\rm s}|\mathcal{H}_1 \sim \mathcal{CN}(0, \kappa)$ with
	$$
	\kappa = |\alpha {\bf v}^H {\bf g}_{\rm t} \left( \{ x_{{\rm R}, n}   \}  \right) {\bf h}_{\rm t}^H \left( \{ x_{{\rm T}, m}   \}  \right) {\bf w} |^2  + \sigma_{\rm s}^2.
	$$
	
	The Neyman-Pearson detector \cite{npd} is employed to judge whether there exists a target, which is formulated as
	\begin{flalign}
		T = |\widetilde{y}_{\rm s}|^2     \underset{\mathcal{H}_0}{\overset{\mathcal{H}_1}{\gtrless}}   \delta, 
	\end{flalign}
	where $\delta$ denotes the decision threshold which is determined by the false alarm probability $P_{\rm FA}$ and the statistic distribution of $T$ is given by
	\begin{subnumcases}
		{T \sim}
		\mathcal{H}_0: \frac{\sigma_{\rm s}^2}{2}  \chi_2^2,                 \\
		\mathcal{H}_1: \frac{\kappa}{2}  \chi_2^2,
	\end{subnumcases}
	where $\chi_2^2$ denotes the central chi-squared distribution with two degree of freedoms. Accordingly, the detection probability $P_{\rm D}$ and false alarm  probability $P_{\rm FA}$ are respectively calculated by
	\begin{flalign}
		& P_{\rm D} = {\rm Pr} \left( T >\delta| \mathcal{H}_1  \right) = 1 - \xi_{\chi_2^2}\left( 2 \delta / \kappa \right), \\
		& P_{\rm FA} = {\rm Pr} \left( T > \delta | \mathcal{H}_0  \right) =  1 - \xi_{\chi_2^2} \left( 2 \delta / \sigma_{\rm s}^2  \right),
	\end{flalign}
	where ${\rm Pr}(\cdot)$ denotes the probability function and $\xi_{\chi_2^2}$ denotes the distribution function of $\chi_2^2$. For a desired $P_{\rm FA}$, the achievable ${P}_{\rm D}$ can be obtained as
	\begin{flalign}
		P_{\rm D} &= 1 - \xi_{\chi_2^2}\left( \frac{\sigma_{\rm s}^2}{\kappa}  \xi_{\chi_2^2}^{-1}(1-P_{\rm FA})    \right) \nonumber \\
		&\propto \frac{\kappa}{\sigma_{\rm s}^2} = 1 + {\gamma}_{\rm t} \left(\{{\bf w}, {\bf v}, x_{{\rm T}, m}, x_{{\rm R}, n} \}  \right), \label{pd_snr}
	\end{flalign}
	where ${\gamma}_{\rm t} \left(\{{\bf w}, {\bf v}, x_{{\rm T}, m}, x_{{\rm R}, n} \}  \right)$ denotes the radar SNR, which is given by
	\begin{flalign}
		{\gamma}_{\rm t} & \left(\{{\bf w}, {\bf v}, x_{{\rm T}, m}, x_{{\rm R}, n} \}  \right) =\nonumber \\
        &\frac{ |\alpha {\bf v}^H {\bf g}_{\rm t} \left( \{ x_{{\rm R}, n}   \}  \right) {\bf h}_{\rm t}^H \left( \{ x_{{\rm T}, m}   \}  \right) {\bf w}|^2}{\sigma_{\rm s}^2 {\bf v}^H {\bf v}} \label{r_snr1} \\
		& \overset{(a)}{ \leq } \frac{ |\alpha|^2 || {\bf g}_{\rm t} \left( \{ x_{{\rm R}, n}   \}  \right)\|^2 |{\bf h}_{\rm t}^H \left( \{ x_{{\rm T}, m}   \}  \right) {\bf w}|^2}{\sigma_{\rm s}^2 },  \label{r_snr2}
	\end{flalign}
	where (a) is due to the Cauchy-Schwarz inequality and the equality holds for ${\bf v} = {\bf g}_t(\{ x_{{\rm R},n} \}) / ||  {\bf g}_t(\{ x_{{\rm R},n} \}) || $. Besides, the RPAs are utilized to received the echo signals reflected by the target, and thus, their locations influence the sensing performance only. Based on the observation, we place RPAs towards the target, i.e., $x_{{\rm R},n} = x_{\rm t}, \forall n \in \mathcal{N} $, to enhance $ ||  {\bf g}_t(\{ x_{{\rm R},n} \}) ||$, and substitute $\{x_{{\rm R},n}\}$ to \eqref{r_snr2} to reach 
    \begin{flalign}
		&\Gamma_{\rm t} \left( \{ {\bf w}, {x}_{{\rm T}, m}   \}  \right) = \\
		&~~~~   \underbrace{ \left(\sum_{n \in \mathcal{N}} \frac{ \eta |\alpha|^2 }{\left(y_{\rm t} - y_{{\rm R}, n}\right)^2 +H^2 } \right)}_{\triangleq \beta} \frac{|{\bf h}_{\rm t}^H( \{ x_{{\rm T},m} \}) {\bf w} |^2}{\sigma_{\rm s}^2},  \nonumber
	\end{flalign}
    which serves an achievable upper bound for the radar SNR (cf. \eqref{r_snr1}) by setting ${\bf v} = {\bf g}_t(\{ x_{{\rm R},n} \}) / ||  {\bf g}_t(\{ x_{{\rm R},n} \}) || $.

	Based on \eqref{pd_snr}, it is observed that the detection probability is positively proportional to the radar SNR. Therefore, we use $\Gamma_{\rm t}(\{ {\bf w}, {x}_{{\rm T}, m}\}) $ as the performance metric for the sensing task.

	\subsection{Problem Formulation}
	
	Our goal is to maximize the communication rate by optimizing the transmit beamforming vector ${\bf w}$ and the TPA locations  $\{ x_{{\rm T}, m} \}$ under the constraints of the transmit power budget, the radar SNR requirement, and the movable area of TPAs. We formulate considered problem as follows:
	\begin{subequations}
		\begin{flalign}
			{{\bf P}_1}: & \mathop{\max}  \limits_{ \{ {\bf w}, x_{{\rm T}, m} \} }  \log_2 \left( 1 + \Gamma_{\rm u} \left(\{ {\bf w}, x_{{\rm T}, m}  \}  \right) \right)  \label{p1a}  \\
			{\rm s.t.} ~&  {\bf w}^H {\bf w} \leq P_{\max},  \label{p1b}  \\
			& \Gamma_{\rm t} \left(\{ {\bf w}, {x}_{{\rm T}, m}  \}  \right)  \geq \Gamma_{\rm Req}, \label{p1c} \\
			& -\frac{L}{2} \leq x_{{\rm T}, m} \leq \frac{L}{2},  \forall m \in \mathcal{M},  \label{p1d}
		\end{flalign}
	\end{subequations}
	where $P_{\max}$ in \eqref{p1b} denotes the transmit power budget and $\Gamma_{\rm Req}$ in \eqref{p1c} denotes the radar SNR threshold associated with the required  detection probability. 
	
	Problem ${\bf P}_1$ is non-convex due to deeply coupling between ${\bf w}$ and $\{ x_{{\rm T}, m} \}$ in the objective function \eqref{p1a} and the constraint \eqref{p1c}. Moreover, the complicated functional relationship of $h_{{\rm u}, m} \left( x_{{\rm T}, m} \right)$/$h_{{\rm t}, m} \left( x_{{\rm T}, m} \right)$ with regards to $ x_{{\rm T}, m} $ makes Problem ${\bf P}_1$ challenging to address.

	\section{Joint Optimization of Deployment and Beamforming Design of Pinching Antennas}
	
	This section first reformulates Problem ${\bf P}_1$ via insightful observations from the considered system and then derives an upper bound problem for the reformulated problem via a fine-tuning method. At last, the  upper bound problem is solved by a proposed SCA-based algorithm.
	
	For notation simplicity, we use ${\bf h}_{\rm u}$ to represent ${\bf h}_{\rm u} ( \{ x_{{\rm T}, m}  \} )$ and ${\bf h}_{\rm t}$ to represent ${\bf h}_{\rm t}( \{ x_{{\rm T}, m}  \} )$
	
	\subsection{Problem Reformulation}

We present two propositions characterizing  the optimal solution structure of Problem ${\bf P_1}$ to facilitate equivalent problem reformation.

	\begin{mypro}
		Suppose that Problem ${\bf P}_1$ is feasible. Let ${\bf w}^{\star}$ be the optimal solution to Problem ${\bf P}_1$. Then, constraint \eqref{p1b} is active with ${\bf w}^{\star}$, i.e.,
		\begin{flalign}
			( {\bf w}^{\star})^H {\bf w}^{\star} = P_{\max}.
		\end{flalign}
	\end{mypro}
	\begin{IEEEproof}
		If $( {\bf w}^{\star})^H {\bf w}^{\star} < P_{\max}$, we can always construct $\widehat{\bf w}^{\star}$ with greater objective result as follows:
		\begin{flalign}
			\widehat{\bf w}^{\star} =  {\bf w}^{\star} + e^{j \theta} \left( \sqrt{P_{\max}} - || {\bf w}^{\star} ||  \right) \frac{{\bf z}}{|| {\bf z} ||},  \label{construct_pmax}
		\end{flalign}
		where $\theta \triangleq {\rm arg} ({\bf h}_{\rm u}^H  {\bf w}^{\star} )$, and ${\bf z}$ is the projection of ${\bf h}_{\rm u}$ onto the space $\Omega_{\rm t} \triangleq \{ {\bf x} \in \mathbb{C}^M | {\bf x}^H {\bf h}_{\rm t} = 0   \}$. 
		
		We can find that 
		\begin{flalign}
			&\left|{\bf h}_{\rm u}^H \widehat{\bf w}^{\star} \right| = \left|{\bf h}_{\rm u}^H   {\bf w}^{\star} + e^{j \theta} \left( \sqrt{P_{\max}} - || {\bf w}^{\star} ||  \right) \frac{{\bf h}_{\rm u}^H {\bf z}}{|| {\bf z} ||}   \right| \nonumber \\
			&~~ = \left|{\bf h}_{\rm u}^H   {\bf w}^{\star}  \right| +  \left( \sqrt{P_{\max}} - || {\bf w}^{\star} ||  \right) {|| {\bf z} ||}  > \left|{\bf h}_{\rm u}^H   {\bf w}^{\star}  \right|, \label{objective} \\ 
			& \left| \left| \widehat{\bf w}^{\star}   \right|    \right| = \left|\left|   {\bf w}^{\star} + e^{j \theta} \left( \sqrt{P_{\max}} - || {\bf w}^{\star} ||  \right) \frac{{\bf z}}{|| {\bf z} ||}      \right|  \right| \nonumber \\
			&~~  \leq \left|\left|   {\bf w}^{\star} \right|  \right| +    \sqrt{P_{\max}} - || {\bf w}^{\star} ||  = \sqrt{P_{\max}}, \label{power_con}    
			\\
			& \left|{\bf h}_{\rm t}^H  \widehat{\bf w}^{\star}  \right| =  \left|{\bf h}_{\rm t}^H   {\bf w}^{\star} + e^{j \theta} \left( \sqrt{P_{\max}} - || {\bf w}^{\star} ||  \right) \frac{{\bf h}_{\rm t}^H {\bf z}}{|| {\bf z} ||}   \right|  \nonumber \\
			&~~ = \left|{\bf h}_{\rm t}^H  {\bf w}^{\star}  \right|.  \label{sen_con} 
		\end{flalign}
		According to \eqref{power_con} and \eqref{sen_con}, $\widehat{\bf w}^{\star}$ satisfies constraint \eqref{p1b} and \eqref{p1c}, which indicates that $\widehat{\bf w}^{\star}$ is feasible to Problem ${\bf P}_1$. Besides, \eqref{objective} implies that   $\widehat{\bf w}^{\star}$ can achieve a higher  communication rate than  ${\bf w}^{\star}$. As a result,  ${\bf w}^{\star}$ is not regarded as the optimal solution, which contradicts the assumption that ${\bf w}^{\star}$ is the optimal solution. Therefore, constraint \eqref{p1b} should be activated with the optimal beamformer. This completes the proof.
	\end{IEEEproof}

	
	\begin{mypro}
		Suppose that Problem ${\bf P}_1$ is feasible. Let ${\bf w}^{\star}$ be the optimal solution to Problem ${\bf P}_1$, ${\bf w}^{\star}$ lies in the space spanned by $\{ {\bf h}_{\rm u}, {\bf h}_{\rm t} \}$, i.e., there exist $c_{\rm u} \in \mathbb{C}$ and $c_{\rm t} \in \mathbb{C}$ such that
		\begin{flalign}
			{\bf w}^{\star} = c_{\rm u} {\bf h}_{\rm u} + c_{\rm t} {\bf h}_{\rm t}.
		\end{flalign}
	\end{mypro}
	\begin{IEEEproof}
		We denote $\Omega_1$ as the space spanned by $\{ {\bf h}_{\rm u}, {\bf h}_{\rm t} \}$  and $\Omega_2 \triangleq {\bf I}_{M} - \Omega_1$ as the orthogonal complementary space of $\Omega_1$. We can find ${\bf x} \in \Omega_1$ and ${\bf y} \in \Omega_2$, which satisfy
		\begin{flalign}
			{\bf w}^{\star} = {\bf x} + {\bf y}. \label{div}
		\end{flalign}
	Then, we construct $\widehat{\bf w}^{\star} = {\bf x}$. If ${\bf y} \neq {\bf 0}_{M}$, one can find $\widehat{\bf w}^{\star}$ satisfying 
		\begin{flalign}
			&|\widehat{\bf w}^{\star} | < |{\bf w}^{\star}| = P_{\max}, \label{power2} \\
			& |{\bf h}_{\rm u}^H {\bf w}^{\star}| =  |{\bf h}_{\rm u}^H {\bf x} + {\bf h}_{\rm u}^H {\bf y} |  = |{\bf h}_{\rm u}^H \widehat{\bf w}^{\star}|, \label{com2}\\
			& |{\bf h}_{\rm t}^H {\bf w}^{\star}| =  |{\bf h}_{\rm t}^H {\bf x} + {\bf h}_{\rm t}^H {\bf y} |  = |{\bf h}_{\rm t}^H \widehat{\bf w}^{\star}|. \label{sen2}
		\end{flalign}
		Combing \eqref{power2}, \eqref{com2}, \eqref{sen2}, and the assumption that ${\bf w}^{\star}$ is optimal, we derive that $\widehat{\bf w}^{\star}$ is another optimal solution to Problem ${\bf P}_1$ with $(\widehat{\bf w}^{\star})^H \widehat{\bf w}^{\star} < P_{\max}$, which is contradictory with Proposition 1.
		
		Therefore, it holds that ${\bf y} = {\bf 0}_{M}$ in \eqref{div}, which indicates that ${\bf w}^{\star} \in \Omega_1$. We  complete the proof.
	\end{IEEEproof}
	
 Proposition 2 re-expresses ${\bf w}$ as  a function of $c_{\rm u}$ and $c_{\rm t}$:
	\begin{flalign}
		{\bf w} = c_{\rm u} {\bf h}_{\rm u} + c_{\rm t} {\bf h}_{\rm t}. \label{re}
	\end{flalign}
	By substituting \eqref{re} into \eqref{p1a}, \eqref{p1b}, and \eqref{p1c}, Problem ${\bf P}_1$ can be rewritten as 
	\begin{subequations}
		\begin{flalign}
			{{\bf P}_2}: & \mathop{\max}  \limits_{ \{ c_{\rm u}\in{\mathbb C}, c_{\rm t}\in{\mathbb C}, x_{{\rm T}, m} \} }  \log_2 \left( 1 +  \frac{f_{\rm u}\left( c_{\rm u}, c_{\rm t}, \{ x_{{\rm T}, m} \}  \right)   }{\sigma_{\rm u}^2} \right)  \label{p2a}  \\
			{\rm s.t.} ~&   f_{\rm p} \left( c_{\rm u}, c_{\rm t}, \{ x_{{\rm T}, m}\}   \right)  \leq P_{\max},  \label{p2b}  \\
			& f_{\rm t} \left( c_{\rm u}, c_{\rm t}, \{ x_{{\rm T}, m}\}   \right)  \geq \frac{\Gamma_{\rm Req} \sigma_{\rm s}^2}{\beta}, \label{p2c} \\
			&\eqref{p1d},  \nonumber
		\end{flalign}
	\end{subequations}
	where the functions $f_{\rm u}( c_{\rm u}, c_{\rm t}, \{x_{{\rm T}, m}\} )$, $f_{\rm p}( c_{\rm u}, c_{\rm t}, \{x_{{\rm T}, m}\} )$, and $f_{\rm t}( c_{\rm u}, c_{\rm t}, \{x_{{\rm T}, m}\} )$ are given by \eqref{f_u}, \eqref{f_p}, and \eqref{f_t}, respectively. 
	\begin{figure*}
		\begin{flalign}
			& f_{\rm u}\left( c_{\rm u}, c_{\rm t}, \{ x_{{\rm T}, m}\}   \right) = \left| c_{\rm u} {\bf h}_{\rm u}^H {\bf h}_{\rm u} + c_{\rm t} {\bf h}_{\rm u}^H {\bf h}_{\rm t}  \right|^2  =  \left| c_{\rm u} \right|^2 \left|\left| {\bf h}_{\rm u} \right| \right|^4 + |c_{\rm t}|^2 |{\bf h}_{\rm t}^H {\bf h}_{\rm u} |^2  + 2 ||{\bf h}_{\rm u} ||^2 {\rm Re} \left\{  c_{\rm u} c_{\rm t}^{*} {\bf h}_{\rm t}^H {\bf h}_{\rm u}  \right\}
			\label{f_u} \\
			&f_{\rm p}\left( c_{\rm u}, c_{\rm t}, \{x_{{\rm T}, m}\}   \right) = \left(  c_{\rm u} {\bf h}_{\rm u} + c_{\rm t} {\bf h}_{\rm t}  \right)^H \left( c_{\rm u} {\bf h}_{\rm u} + c_{\rm t} {\bf h}_{\rm t}\right) = |c_{\rm u}|^2 ||{\bf h}_{\rm u} ||^2 + |c_{\rm t}|^2 ||{\bf h}_{\rm t} ||^2 + 2 {\rm Re} \left\{ c_{\rm u} c_{\rm t}^{*} {\bf h}_{\rm t}^H {\bf h}_{\rm u}    \right\}  \label{f_p} \\
			& f_{\rm t}\left( c_{\rm u}, c_{\rm t}, \{x_{{\rm T}, m}\}   \right) = \left|c_{\rm u} {\bf h}_{\rm t}^H {\bf h}_{\rm u} + c_{\rm t} {\bf h}_{\rm t}^H {\bf h}_{\rm t}   \right|^2 = |c_{\rm u}|^2 |{\bf h}_{\rm t}^H {\bf h}_{\rm u} |^2 + |c_{\rm t}|^2 ||{\bf h}_{\rm t}||^4 + 2  ||{\bf h}_{\rm t}||^2 {\rm Re}\left\{ c_{\rm u} c_{\rm t}^{*} {\bf h}_{\rm t}^H {\bf h}_{\rm u}   \right\}
			\label{f_t}
		\end{flalign}
		\hrule
	\end{figure*}
	
	\subsection{Problem Approximation}

It is challenging to directly handle Problem ${\bf P}_2$, since the TPA locations, i.e., $\{ x_{{\rm T}, m} \}$, affect both large-scale path loss and small-scale phase shift. Fortunately, the following two observations are critical for simplifying the considered problem. First, the effect of wavelength-scale TPA location adjustments on path loss is negligible \cite{pinch}. Second, the correlation between communication and sensing channels, i.e., $|{\bf h}_{\rm t}^H {\bf h}_{\rm u}| / (||{\bf h}_{\rm t}|| \times ||{\bf h}_{\rm u}||)$, has a significant impact on the communication-sensing performance trade-off. Therefore, we separate the influence of TPA locations on the path loss and phase shift by fine tuning $\{ x_{{\rm T}, m} \}$ to maximize $|{\bf h}_{\rm t}^H {\bf h}_{\rm u}|$. We prove that this fine-tuning method has negligible performance loss on the system performance, which is summarized in Proposition 3.
    
	\begin{mypro}
		Let $\{ c_{\rm u}^{\star},   c_{\rm t}^{\star}, x_{{\rm T}, m}^{\star} \}$ be the optimal solution to Problem ${\bf P}_2$, and         ${\bf h}_{\rm t}^{\star}$/${\bf h}_{\rm u}^{\star}$ and $\widehat{\bf h}_{\rm t}^{\star}$/$\widehat{\bf h}_{\rm u}^{\star}$ be the channels before and after fine tuning, respectively.
		
		We can construct another solution to Problem  ${\bf P}_2$ denoted by $\{\widehat{c}^{\star}_{\rm u}, \widehat{c}^{\star}_{\rm t}, \widehat{x}_{{\rm T}, m}^{\star}\}$ by fine tuning  $\{ c_{\rm u}^{\star},   c_{\rm t}^{\star}, x_{{\rm T}, m}^{\star} \}$ such that
		\begin{flalign}
			&\theta_m \left( \widehat{x}_{{\rm T}, m}^{\star} \right) \triangleq \frac{\left|  \left| {\bm \psi}_{\rm t} -  \widehat{\bm \psi}_{{\rm T}, m}^{\star}  \right| \right|}{\lambda}   -  \frac{\left|  \left| {\bm \psi}_{\rm u} -  \widehat{\bm \psi}_{{\rm T}, m}^{\star}  \right| \right|}{\lambda} \in \mathbb{N},  \label{re_theta}\\
			&\widehat{c}^{\star}_{\rm u} = {c}^{\star}_{\rm u} e^{j \zeta},\\
			&{\rm and}~\widehat{c}^{\star}_{\rm t} = {c}^{\star}_{\rm t},
		\end{flalign}
		with $\zeta$ being the solution to the follow equation:
		\begin{flalign}
			{\rm Re} \left\{ e^{j\zeta}  c_{\rm u}^{\star}  \left( c_{\rm t}^{\star}  \right)^{*} \left(\widehat{\bf h}_{\rm t}^{\star} \right)^H \widehat{\bf h}_{\rm u}^{\star} -  c_{\rm u}^{\star}  \left( c_{\rm t}^{\star}  \right)^{*} \left({\bf h}_{\rm t}^{\star} \right)^H {\bf h}_{\rm u}^{\star}       \right\} = 0. \label{zeta}
		\end{flalign}
		
		Fine tuning $\{x^{\star}_{{\rm T},m}\}$ to $\{\widehat{x}^{\star}_{{\rm T},m}\}$ to satisfy \eqref{re_theta} can be within a few wavelength such that  $\|{\bf h}_{\rm t}^{\star}\|\approx\|\widehat{\bf h}_{\rm t}^{\star}\|$ and $\|{\bf h}_{\rm u}^{\star}\|\approx\|\widehat{\bf h}_{\rm u}^{\star}\|$, and the performance loss due to fine tuning is on the order of the wavelength.
	\end{mypro}
    
	\begin{IEEEproof}
		Please refer to Appendix A.
	\end{IEEEproof}

	Based on Proposition 3, an upper-bound problem of Problem ${\bf P}_2$ is obtained as follows:
	\begin{subequations}
		\begin{flalign}
			{{\bf P}_3}: & \mathop{\max}  \limits_{ \{ c_{\rm u}\in{\mathbb C}, c_{\rm t}\in{\mathbb C}, x_{{\rm T}, m} \} }  \log_2 \left( 1 +  \frac{\widehat{f}_{\rm u}\left( c_{\rm u}, c_{\rm t}, \{ x_{{\rm T}, m}\}   \right)   }{\sigma_{\rm u}^2} \right)  \label{p3a}  \\
			{\rm s.t.} ~&   \widehat{f}_{\rm p} \left( c_{\rm u}, c_{\rm t}, \{x_{{\rm T}, m}\}   \right)  \leq P_{\max},  \label{p3b}  \\
			& \widehat{f}_{\rm t} \left( c_{\rm u}, c_{\rm t}, \{x_{{\rm T}, m}\}   \right)  \geq \frac{\Gamma_{\rm Req} \sigma_{\rm s}^2}{\beta}, \label{p3c} \\
			&\eqref{p1d},  \nonumber
		\end{flalign}
	\end{subequations}
	where the functions  $\widehat{f}_{\rm u}\left( c_{\rm u}, c_{\rm t}, \{x_{{\rm T}, m}\}   \right) $, $\widehat{f}_{\rm p}\left( c_{\rm u}, c_{\rm t}, \{x_{{\rm T}, m}\}   \right) $, and $\widehat{f}_{\rm t}\left( c_{\rm u}, c_{\rm t}, \{x_{{\rm T}, m}\}   \right) $ are respectively given by \eqref{hat_f_u}, \eqref{hat_f_p}, and \eqref{hat_f_t} with 
	\begin{flalign*}
		& d_{{\rm u}, m}^2  = (y_{\rm u} - y_{{\rm T}, m}  )^2 + H^2 ~{\rm and}~d_{{\rm t}, m}^2  = (y_{\rm t} - y_{{\rm T}, m}  )^2 + H^2. 
	\end{flalign*}
	
	\begin{figure*}
		\begin{flalign}
			&\widehat{f}_{\rm u}\left( c_{\rm u}, c_{\rm t}, \{x_{{\rm T}, m}\}   \right) =  \left(\sum_{m \in \mathcal{M}} \frac{\eta |c_{\rm u}|}{\left( x_{\rm u} - x_{{\rm T}, m} \right)^2 + d_{{\rm u}, m}^2  } \right)^2 +  \left| \sum_{m \in \mathcal{M}} \frac{\eta |c_{\rm t}| }{\sqrt{ \left( \left(x_{\rm u} - x_{{\rm T}, m}\right)^2  + d_{{\rm u}, m}^2 \right) \left( \left(x_{\rm t} - x_{{\rm T}, m}\right)^2  + d_{{\rm t}, m}^2 \right)   }}  \right|^2 
			\nonumber \\
			&~  +
			\left(\sum_{m \in \mathcal{M}} \frac{\eta}{\left( x_{\rm u} - x_{{\rm T}, m} \right)^2 + d_{{\rm u}, m}^2  } \right) \left( \sum_{m \in \mathcal{M}} \frac{   \eta \left( |c_{\rm u} + c_{\rm t}|^2 - |c_{\rm u}|^2 - |c_{\rm t}|^2   \right) }{\sqrt{\left( \left(x_{\rm u} -x_{{\rm T}, m}\right)^2 + d_{{\rm u}, m}^2 \right) \left( \left(x_{\rm t} -x_{{\rm T}, m}\right)^2 + d_{{\rm t}, m}^2 \right)     }   }  \right)  
			\label{hat_f_u} \\
			&\widehat{f}_{\rm p}\left( c_{\rm u}, c_{\rm t}, \{x_{{\rm T}, m}\}   \right) =   \label{hat_f_p} \\
			&~   \sum_{m \in \mathcal{M}} \left(\frac{\eta |c_{\rm u}|^2 }{\left(x_{\rm u} - x_{{\rm T}, m}  \right)^2 + d_{{\rm u}, m}^2 }   +   \frac{\eta |c_{\rm t}|^2}{\left(x_{\rm t} - x_{{\rm T}, m}  \right)^2 + d_{{\rm t}, m}^2} +   \frac{  \eta \left( |c_{\rm u} + c_{\rm t}|^2 - |c_{\rm u}|^2 - |c_{\rm t}|^2   \right) }{\sqrt{ \left( \left(x_{\rm u} - x_{{\rm T}, m}\right)^2  + d_{{\rm u}, m}^2 \right) \left( \left(x_{\rm t} - x_{{\rm T}, m}\right)^2  + d_{{\rm t}, m}^2 \right)   }}    \right) \nonumber \\
			& \widehat{f}_{\rm t}\left( c_{\rm u}, c_{\rm t}, \{x_{{\rm T}, m}\}   \right) = 
			  \left| \sum_{m \in \mathcal{M}} \frac{\eta |c_{\rm u}| }{\sqrt{ \left( \left(x_{\rm u} - x_{{\rm T}, m}\right)^2  + d_{{\rm u}, m}^2 \right) \left( \left(x_{\rm t} - x_{{\rm T}, m}\right)^2  + d_{{\rm t}, m}^2 \right)   }}  \right|^2  
			+ 
			 \left( \sum_{m \in \mathcal{M}} \frac{\eta |c_{\rm t}|}{\left(x_{\rm t} - x_{{\rm T}, m} \right)^2 + d_{{\rm t}, m}^2} \right)^2  \nonumber \\
			&~+ 
			\left( \sum_{m \in \mathcal{M}} \frac{\eta}{\left(x_{\rm t} - x_{{\rm T}, m} \right)^2 + d_{{\rm t}, m}^2}    \right)
			\left( \sum_{m \in \mathcal{M}} \frac{ \eta \left( |c_{\rm u} + c_{\rm t}|^2 - |c_{\rm u}|^2 - |c_{\rm t}|^2   \right)  }{\sqrt{ \left( \left(x_{\rm u} - x_{{\rm T}, m}\right)^2  + d_{{\rm u}, m}^2 \right) \left( \left(x_{\rm t} - x_{{\rm T}, m}\right)^2  + d_{{\rm t}, m}^2 \right)   }}    \right) \label{hat_f_t}
		\end{flalign}
		\hrule
	\end{figure*}
	
	Problem ${\bf P}_3$ is non-convex due to the objective function \eqref{p3a} and the constraints \eqref{p3b} and \eqref{p3c}. To handle the complicated coupling among $c_{\rm u}$, $c_{\rm t}$, and $\{x_{{\rm T}, m}\}$, auxiliary variables $\{a_{{\rm u}, m} \}$, $\{b_{{\rm u}, m}\}$, $\{a_{{\rm t}, m} \}$, and $\{b_{{\rm t}, m}\}$ are introduced, and they satisfy 
	\begin{flalign}
		&\left(x_{\rm u} - x_{{\rm T}, m} \right)^2 + d_{{\rm u}, m}^2 \geq e^{-2 a_{{\rm u},m}}, \label{aum} \\
		&\left(x_{\rm u} - x_{{\rm T}, m} \right)^2 + d_{{\rm u}, m}^2 \leq e^{-2 b_{{\rm u}, m}}, \label{bum} \\
		&\left(x_{\rm t} - x_{{\rm T}, m} \right)^2 + d_{{\rm t}, m}^2 \geq e^{-2 a_{{\rm t},m}}, \label{atm}  \\
		&\left(x_{\rm t} - x_{{\rm T}, m} \right)^2 + d_{{\rm t}, m}^2 \leq e^{-2 b_{{\rm t}, m}}. \label{btm}
	\end{flalign}
	Then, \eqref{p3a}, \eqref{p3b}, and \eqref{p3c} are respectively rewritten as
	\begin{flalign}
		& \log_2\left(1 + \frac{\eta^2 g_{\rm u}({\bf L}_1) }{\sigma_{\rm u}^2}\right),  \label{p3a1}  \\
		& g_{\rm p}({\bf L}_1) \leq \frac{P_{\max}}{\eta},  \label{p3b1}  \\
		& g_{\rm t} ({\bf L}_1) \geq \frac{\Gamma_{\rm Req} \sigma_{\rm s}^2}{\beta \eta^2},  \label{p3c1}
	\end{flalign}
	where ${\bf L}_1 \triangleq \{c_{\rm u}, c_{\rm t}, x_{{\rm T}, m}, a_{{\rm u}, m}, b_{{\rm u}, m}, a_{{\rm t}, m}, b_{{\rm t}, m}  \}$ and $g_{\rm u}({\bf L}_1)$, $g_{\rm p}({\bf L}_1)$, and $g_{\rm t}({\bf L}_1)$ are given by \eqref{gu}, \eqref{gp}, and \eqref{gt}, respectively.   
	\begin{figure*}
		\begin{flalign}
			&  g_{\rm u} \left( {\bf L}_1 \right) =  |c_{\rm u}|^2 \left( \sum_{m \in \mathcal{M}} e^{2 b_{{\rm u}, m}}\right)^2 + |c_{\rm t}|^2 \left( \sum_{m \in \mathcal{M}} e^{b_{{\rm u}, m} + b_{{\rm t}, m}}\right)^2 +  |c_{\rm u}+ c_{\rm t} |^2 \left( \sum_{m \in \mathcal{M}} e^{2 b_{{\rm u}, m}}  \right)\left( \sum_{m \in \mathcal{M}} e^{b_{{\rm u}, m}+b_{{\rm t}, m}} \right) \nonumber \\
			&~ - (|c_{\rm u}|^2 + |c_{\rm t}|^2) \left(\sum_{m \in \mathcal{M}} e^{2 a_{{\rm u}, m}}   \right)\left(\sum_{m \in \mathcal{M}} e^{a_{{\rm u}, m} + a_{{\rm t}, m}  }   \right) \label{gu}  \\
			& g_{\rm p} \left( {\bf L}_1 \right) = \sum_{m \in \mathcal{M}} \left( |c_{\rm u}|^2 e^{2 a_{{\rm u}, m}} + |c_{\rm t}|^2 e^{2 a_{{\rm t}, m}} + |c_{\rm u}+ c_{\rm t} |^2 e^{a_{{\rm u}, m} + a_{{\rm t}, m}} - (|c_{\rm u}|^2 + |c_{\rm t}|^2) e^{b_{{\rm u}, m}+ b_{{\rm t}, m}} \right) \label{gp} \\
			& g_{\rm t} \left({\bf L}_1 \right) = |c_{\rm u}|^2 \left(\sum_{m \in \mathcal{M}} e^{b_{{\rm u}, m} + b_{{\rm t}, m}}  \right)^2 + |c_{\rm t}|^2 \left(\sum_{m \in \mathcal{M}} e^{2 b_{{\rm t}, m} }  \right)^2 + |c_{\rm u} + c_{\rm t}|^2 \left( \sum_{m \in \mathcal{M}}  e^{2 b_{{\rm t},m}}  \right) \left( \sum_{m \in \mathcal{M}} e^{b_{{\rm u}, m} + b_{{\rm t}, m}} \right) \nonumber \\
			& - ( |c_{\rm u}|^2 + |c_{\rm t}|^2 )\left(\sum_{m \in \mathcal{M}} e^{2 a_{{\rm t}, m} } \right) \left( \sum_{m \in \mathcal{M}} e^{a_{{\rm u}, m} + a_{{\rm t}, m} }  \right) \label{gt}
		\end{flalign}
		\hrule
	\end{figure*}
	To address the coupling between $\{c_{\rm u}, c_{\rm t}\}$ and $\{a_{{\rm u}, m}, b_{{\rm u}, m}, a_{{\rm t}, m}, b_{{\rm t}, m} \}$, the auxiliary optimization variables $p_{\rm u}$, $q_{\rm u}$, $p_{\rm t}$, $q_{\rm t}$, $o$, and $v$ are introduced and satisfy
	\begin{flalign}
		& e^{p_{\rm u}} \leq |c_{\rm u}|^2 \leq e^{q_{\rm u}}, \label{pqu} \\
		& e^{p_{\rm t}} \leq |c_{\rm t}|^2 \leq e^{q_{\rm t}}, \label{pqt} \\
		& e^{o} \leq |c_{\rm u}+c_{\rm t}|^2 \leq e^{v}. \label{ov}  
	\end{flalign}
	With the introduced $\{ p_{\rm u}, q_{\rm u}, p_{\rm t}, q_{\rm t}, o, v \}$, \eqref{p3a1}, \eqref{p3b1}, and \eqref{p3c1} are respectively re-expressed as 
	\begin{flalign}
		& \log_2\left(1 + \frac{\eta^2 \widehat{g}_{\rm u}({\bf L}_2) }{\sigma_{\rm u}^2}\right),  \label{p3a2}  \\
		& \widehat{g}_{\rm p}({\bf L}_2) \leq \frac{P_{\max}}{\eta},  \label{p3b2}  \\
		& \widehat{g}_{\rm t} ({\bf L}_2) \geq \frac{\Gamma_{\rm Req} \sigma_{\rm s}^2}{\beta \eta^2},  \label{p3c2}
	\end{flalign}
	where ${\bf L}_2 \triangleq {\bf L}_1 \cup \{p_{\rm u}, q_{\rm u}, p_{\rm t}, q_{\rm t}, o, v\}$ and $\widehat{g}_{\rm u}({\bf L}_2)$, $\widehat{g}_{\rm p}({\bf L}_2)$, and $\widehat{g}_{\rm t}({\bf L}_2)$ are given by \eqref{hat_gu}, \eqref{hat_gp}, and \eqref{hat_gt}, respectively.
	\begin{figure*}
		\begin{flalign}
			&\widehat{g}_{\rm u} ({\bf L}_2) =  e^{p_{\rm u}} \left( \sum_{m \in \mathcal{M}} e^{2 b_{{\rm u}, m}}\right)^2 + e^{p_{\rm t}} \left( \sum_{m \in \mathcal{M}} e^{b_{{\rm u}, m} + b_{{\rm t}, m}}\right)^2 +  e^{o} \left( \sum_{m \in \mathcal{M}} e^{2 b_{{\rm u}, m}}  \right)\left( \sum_{m \in \mathcal{M}} e^{b_{{\rm u}, m}+b_{{\rm t}, m}} \right) \nonumber \\
			&~ - (e^{q_{\rm u}} + e^{q_{\rm t}}) \left(\sum_{m \in \mathcal{M}} e^{2 a_{{\rm u}, m}}   \right)\left(\sum_{m \in \mathcal{M}} e^{a_{{\rm u}, m} + a_{{\rm t}, m}  }   \right) \label{hat_gu}  \\
			& \widehat{g}_{\rm p} ({\bf L}_2) = \sum_{m \in \mathcal{M}} \left( e^{2 a_{{\rm u}, m} + q_{\rm u}} +  e^{2 a_{{\rm t}, m} + q_{\rm t}} +  e^{a_{{\rm u}, m} + a_{{\rm t}, m} + v} - (e^{p_{\rm u}} + e^{p_{\rm t}}) e^{b_{{\rm u}, m}+ b_{{\rm t}, m}} \right) \label{hat_gp} \\
			& \widehat{g}_{\rm t} ({\bf L}_2) = e^{p_{\rm u}} \left(\sum_{m \in \mathcal{M}} e^{b_{{\rm u}, m} + b_{{\rm t}, m}}  \right)^2 + e^{p_{\rm t}} \left(\sum_{m \in \mathcal{M}} e^{2 b_{{\rm t}, m} }  \right)^2 + e^o \left( \sum_{m \in \mathcal{M}}  e^{2 b_{{\rm t},m}}  \right) \left( \sum_{m \in \mathcal{M}} e^{b_{{\rm u}, m} + b_{{\rm t}, m}} \right) \nonumber \\
			& -  (e^{q_{\rm u}} + e^{q_{\rm t}})\left(\sum_{m \in \mathcal{M}} e^{2 a_{{\rm t}, m} } \right) \left( \sum_{m \in \mathcal{M}} e^{a_{{\rm u}, m} + a_{{\rm t}, m} }  \right) \label{hat_gt}
		\end{flalign}
		\hrule
	\end{figure*}
	
	After the aforementioned reformulations, Problem ${\bf P}_3$ is rewritten as
	\begin{subequations}
		\begin{flalign}
			{{\bf P}_4}: & \mathop{\max}  \limits_{ {\bf L}_2 }~  \widehat{g}_{\rm u} ({\bf L}_2) \label{p4a}  \\
			{\rm s.t.} ~
			&\eqref{p1d}, \eqref{aum}, \eqref{bum}, \eqref{atm}, \eqref{btm}, \eqref{pqu}, \eqref{pqt}, \eqref{ov}, \eqref{p3b2}, \eqref{p3c2}.  \nonumber
		\end{flalign}
	\end{subequations}
	Problem ${\bf P}_4$ is non-convex due to the objective function \eqref{p4a} and constraints \eqref{aum}, \eqref{bum}, \eqref{atm}, \eqref{btm}, \eqref{pqu}, \eqref{pqt}, \eqref{ov}, \eqref{p3b2}, and \eqref{p3c2}. Nevertheless, these constraints can be approximated by convex functions based on the first-order Taylor approximation, which are respectively given by
	\begin{flalign}
		& \widetilde{g}_{\rm u}({\bf L}_2, \widetilde{\bf L}_2  ),  \\
		& \left(x_{\rm u} - \widetilde{x}_{{\rm T}, m} \right)^2 + 2 \left(\widetilde{x}_{{\rm T}, m} - x_{\rm u}   \right) \left( x_{{\rm T}, m} - \widetilde{x}_{{\rm T}, m}  \right) + d_{{\rm u}, m}^2 \nonumber \\
		&~~~~ \geq e^{-2 a_{{\rm u},m}}, \label{tilde_aum} \\
		& \left(x_{\rm u} - x_{{\rm T}, m} \right)^2 + d_{{\rm u}, m}^2 \leq e^{-2 \widetilde{b}_{{\rm u}, m}} \left(1 - 2 \left(b_{{\rm u}, m} - \widetilde{b}_{{\rm u}, m} \right)\right) , \label{tilde_bum} \\
		&  \left(x_{\rm t} - \widetilde{x}_{{\rm T}, m} \right)^2 + 2 \left(\widetilde{x}_{{\rm T}, m} -x_{\rm t}   \right) \left(x_{{\rm T}, m} - \widetilde{x}_{{\rm T}, m}  \right) + d_{{\rm t}, m}^2 \nonumber \\
		&~~~~ \geq e^{-2 a_{{\rm t},m}}, \label{tilde_atm}  \\
		& \left(x_{\rm t} - x_{{\rm T}, m} \right)^2 + d_{{\rm t}, m}^2 \leq e^{-2 \widetilde{b}_{{\rm t}, m}}\left(1 - 2\left(b_{{\rm t}, m} - \widetilde{b}_{{\rm t}, m}\right)  \right), \label{tilde_btm} \\
		& e^{p_{\rm u}} \leq  2 {\rm Re}\left\{\widetilde{c}_{\rm u}^{*} c_{\rm u}  \right\} - |\widetilde{c}_{\rm u}|^2,~ |c_{\rm u}|^2 \leq e^{\widetilde{q}_{\rm u}}\left(1 + q_{\rm u} - \widetilde{q}_{\rm u}\right), \label{tilde_pqu} \\
		& e^{p_{\rm t}} \leq  2 {\rm Re}\left\{ \widetilde{c}_{\rm t}^{*}  c_{\rm t}  \right\} - |\widetilde{c}_{\rm t}|^2 , ~  |c_{\rm t}|^2 \leq e^{\widetilde{q}_{\rm t}} \left(1 + q_{\rm t} - \widetilde{q}_{\rm t}\right), \label{tilde_pqt} \\
		& e^{o} \leq  2 {\rm Re}\left\{ (\widetilde{c}_{\rm u}^{*}+\widetilde{c}_{\rm t}^{*} ) (c_{\rm u} + c_{\rm t})    \right\} - |\widetilde{c}_{\rm u}+\widetilde{c}_{\rm t}|^2 , \nonumber \\
		&|c_{\rm u}+c_{\rm t}|^2 \leq e^{\widetilde{v}}\left( 1 + v - \widetilde{v}  \right), \label{tilde_ov} \\
		& \widetilde{g}_{\rm p} \left({\bf L}_2, \widetilde{\bf L}_2  \right) \leq \frac{P_{\max}}{\eta},  \label{p3b_tilde}  \\
		& \widetilde{g}_{\rm t} \left( {\bf L}_2, \widetilde{\bf L}_2 \right)\geq \frac{\Gamma_{\rm Req} \sigma_{\rm s}^2}{\beta \eta^2},  \label{p3c_tilde}
	\end{flalign}
	where 
	$$\widetilde{\bf L}_2 \triangleq \{\widetilde{c}_{\rm u}, \widetilde{c}_{\rm t}, \widetilde{x}_{{\rm T}, m}, \widetilde{a}_{{\rm u}, m}, \widetilde{b}_{{\rm u}, m}, \widetilde{a}_{{\rm t}, m}, \widetilde{b}_{{\rm t}, m}, \widetilde{p}_{\rm u}, \widetilde{q}_{\rm u}, \widetilde{p}_{\rm t}, \widetilde{q}_{\rm t}, \widetilde{o}, \widetilde{v} \}$$
	is a feasible solution to Problem ${\bf P}_4$ and $\widetilde{g}_{\rm u} ({\bf L}_2, \widetilde{\bf L}_2  )$, $\widetilde{g}_{\rm p} ({\bf L}_2, \widetilde{\bf L}_2  )$, and $\widetilde{g}_{\rm t} ({\bf L}_2, \widetilde{\bf L}_2  )$ are given by \eqref{gu_tilde}, \eqref{gp_tilde}, and \eqref{gt_tilde}, respectively.
	\begin{figure*}
		\begin{flalign}
			\widetilde{g}_{\rm u} \left({\bf L}_2, \widetilde{\bf L}_2  \right) = &~  e^{\widetilde{p}_{\rm u}}\left(\sum_{m \in \mathcal{M}} e^{2 \widetilde{b}_{{\rm u}, m}} \right) \left(\sum_{m \in \mathcal{M}} e^{2 \widetilde{b}_{{\rm u}, m}}  \left(1+p_{\rm u} + 4 b_{{\rm u}, m}  - \widetilde{p}_{\rm u} -  4  \widetilde{b}_{{\rm u}, m}    \right) \right) + e^{\widetilde{p}_{\rm t}} \left(\sum_{m \in \mathcal{M}} e^{\widetilde{b}_{{\rm u}, m} + \widetilde{b}_{{\rm t}, m} }  \right)  \times  \nonumber \\
			&  \Bigg(\sum_{m \in \mathcal{M}} e^{\widetilde{b}_{{\rm u}, m} + \widetilde{b}_{{\rm t}, m} }\left( 1 + p_{\rm t} + 2 b_{{\rm u}, m} + 2 b_{{\rm t}, m}  - \widetilde{p}_{\rm t} -  2 \widetilde{b}_{{\rm u}, m} - 2 \widetilde{b}_{{\rm t}, m}  \right) \Bigg)  + e^{\widetilde{o}} \left( \sum_{m \in \mathcal{M}} e^{2\widetilde{b}_{{\rm u}, m}}  \right) \times  \nonumber \\
			& \Bigg(\sum_{m \in \mathcal{M}} e^{\widetilde{b}_{{\rm u}, m} + \widetilde{b}_{{\rm t}, m}}   \left( 1 + o + b_{{\rm u}, m} + b_{{\rm t}, m} - \widetilde{o}- \widetilde{b}_{{\rm u}, m} - \widetilde{b}_{{\rm t}, m}  \right) \Bigg) + e^{\widetilde{o}} \left(\sum_{m \in \mathcal{M}} 2 e^{2 \widetilde{b}_{{\rm u}, m}}\left(b_{{\rm u}, m} - \widetilde{b}_{{\rm u}, m}  \right)   \right) \times \nonumber \\
			& \left(\sum_{m \in \mathcal{M}} e^{\widetilde{b}_{{\rm u}, m} + \widetilde{b}_{{\rm t}, m}   }  \right) - \left(e^{q_{\rm u}} + e^{q_{\rm t}} \right) \left(\sum_{m \in \mathcal{M}} e^{2 a_{{\rm u}, m}}  \right) \left(\sum_{m \in \mathcal{M}} e^{a_{{\rm u}, m} + a_{{\rm t}, m}  }   \right) \label{gu_tilde}  \\
			\widetilde{g}_{\rm p} \left({\bf L}_2, \widetilde{\bf L}_2  \right) = & \sum_{m \in \mathcal{M}} \left( e^{2 a_{{\rm u}, m} + q_{\rm u}} +  e^{2 a_{{\rm t}, m} + q_{\rm t}} +  e^{a_{{\rm u}, m} + a_{{\rm t}, m} + v}  \right) -\sum_{m \in \mathcal{M}} e^{\widetilde{b}_{{\rm u}, m} + \widetilde{b}_{{\rm t}, m}} \left( \left( e^{\widetilde{p}_{\rm u}} + e^{\widetilde{p}_{\rm t}} \right)  \left( 1 + b_{{\rm u}, m} + b_{{\rm t}, m} \right. \right.
			\nonumber \\
			&\left. \left.
			- \widetilde{b}_{{\rm u}, m} - \widetilde{b}_{{\rm t}, m}   \right)  \right)   -\sum_{m \in \mathcal{M}} e^{\widetilde{b}_{{\rm u}, m} + \widetilde{b}_{{\rm t}, m}} \left(  e^{\widetilde{p}_{\rm u}}  \left( p_{\rm u}  - \widetilde{p}_{\rm u} \right) + e^{\widetilde{p}_{\rm t}}\left( p_{\rm t} - \widetilde{p}_{\rm t}  \right)  \right) \label{gp_tilde}  \\
			\widetilde{g}_{\rm t} \left( {\bf L}_2, \widetilde{\bf L}_2 \right) = &~ e^{\widetilde{p}_{\rm u}} \left(\sum_{m \in \mathcal{M}} e^{\widetilde{b}_{{\rm u}, m} + \widetilde{b}_{{\rm t}, m} }  \right) \left(\sum_{m \in \mathcal{M}} e^{\widetilde{b}_{{\rm u}, m} + \widetilde{b}_{{\rm t}, m} } \left( 1 + p_{\rm u} + 2 b_{{\rm u}, m} + 2 b_{{\rm t}, m} - \widetilde{p}_{\rm u} - 2 \widetilde{b}_{{\rm u}, m} - 2 \widetilde{b}_{{\rm t}, m}   \right)  \right)  +  e^{\widetilde{p}_{\rm t}}  \times \nonumber \\
			& \left(\sum_{m \in \mathcal{M}} e^{2 \widetilde{b}_{{\rm t}, m} } \right)  \left(\sum_{m \in \mathcal{M}} e^{2 \widetilde{b}_{{\rm t}, m} } \left(1 + p_{\rm t} +  4 b_{{\rm t}, m} - \widetilde{p}_{\rm t} - 4 \widetilde{b}_{{\rm t}, m} \right)  \right) +   e^{\widetilde{o}} \left( \sum_{m \in \mathcal{M}} e^{2 \widetilde{b}_{{\rm t}, m}}  \right) \Bigg(\sum_{m \in \mathcal{M}} e^{\widetilde{b}_{{\rm u}, m} + \widetilde{b}_{{\rm t}, m}}  \nonumber \\
			&  \left(  1 + o + b_{{\rm u}, m} + b_{{\rm t}, m} - \widetilde{o} - \widetilde{b}_{{\rm u}, m} - \widetilde{b}_{{\rm t}, m} \right)  \Bigg) + e^{\widetilde{o}} \left(\sum_{m \in \mathcal{M}} 2 e^{2 \widetilde{b}_{{\rm t}, m} } \left( b_{{\rm t}, m} -  \widetilde{b}_{{\rm t}, m}  \right) \right) \left(\sum_{m \in \mathcal{M}} e^{\widetilde{b}_{{\rm u}, m} + \widetilde{b}_{{\rm t}, m}} \right) \nonumber\\
			& - \left(e^{q_{\rm u}} + e^{q_{\rm t}} \right) \left(\sum_{m \in \mathcal{M}} e^{2 a_{{\rm t}, m}}  \right) \left(\sum_{m \in \mathcal{M}} e^{a_{{\rm u}, m} + a_{{\rm t}, m}  }   \right) \label{gt_tilde} 
		\end{flalign}
		\hrule
	\end{figure*}
	With a given $\widetilde{\bf L}_2$, Problem ${\bf P}_4$ can be approximated by the following convex optimization problem:
	\begin{subequations}
		\begin{flalign}
			{{\bf P}_5}: & \mathop{\max}  \limits_{ {\bf L}_2 }~  \widetilde{g}_{\rm u} ({\bf L}_2, \widetilde{\bf L}_2) \label{p5a}  \\
			{\rm s.t.} ~
			&\eqref{p1d}, \eqref{tilde_aum}, \eqref{tilde_bum}, \eqref{tilde_atm}, \eqref{tilde_btm}, \eqref{tilde_pqu}, \eqref{tilde_pqt}, \eqref{tilde_ov}, \eqref{p3b_tilde}, \eqref{p3c_tilde}.  \nonumber
		\end{flalign}
	\end{subequations}
	
	Moreover, the SCA method is applied to enhance the approximation precision by iteratively solving Problem ${\bf P}_5$ and updating $\widetilde{\bf L}_2$. 
	
	\section{Simulation Results}
	This section provides simulation results to 
 evaluate the proposed pinching-antenna ISAC system, compared with the conventional ISAC system. Besides, the effectiveness of 
 the proposed algorithm is also demonstrated. In our simulations, unless otherwise specified, the default configuration is detailed as follows \cite{pinch, pa_isac1, pa_isac4}. The length and width of the serving area are set as $L = 40~{\rm m}$ and $W = 20~{\rm m}$, respectively; the height of the waveguide is set as $H = 3~{\rm m}$; the speed of light is $3 \times 10^8~{\rm m/s}$ and the wavelength of carrier frequency is set as $\lambda = 5~{\rm cm}$; the effective refractive index of each dielectric waveguide is set as $n_{\rm eff} = 1.4$; the number of TPAs and RPAs are set as $M = 4$ and $N = 4$, respectively, where the TPAs and RPAs are uniformly distributed within the serving area; the transmit power budget is set as $P_{\max} = 10~{\rm W}$; the radar SNR requirement is set as $\Gamma_{\rm Req} = 4$; the noise power at the user is set as $\sigma_{\rm u}^2 = -60~{\rm dBm}$, and that at the RPAs is set as $\sigma_{\rm s}^2 = -80~{\rm dBm}$.
	
	\subsection{Special Case Analysis}
	In this subsection, we quantitatively analyze the impact of  relative position between the user and the target on the communication performance in the pinching-antenna  ISAC system by considering $3$ special cases with $M=2$. Specifically, for each case, the $1$-st TPA is on the waveguide $y_{{\rm T}, 1} = W/3$ and the $2$-nd TPA is on the waveguide $y_{{\rm T}, 2} = 2W/3$, while the coordinates of user and target in Case1, Case2, and Case3 are set as ${\bm \psi}_{\rm u}^{\rm I} = [L/10, 2W/5, 0]^T$ and  ${\bm \psi}_{\rm t}^{\rm I} = [-L/10, 3W/5, 0]^T$,  ${\bm \psi}_{\rm u}^{\rm II} = [L/5, 2W/5, 0]^T$ and  ${\bm \psi}_{\rm t}^{\rm II} = [-L/5, 3W/5, 0]^T$, and ${\bm \psi}_{\rm u}^{\rm III} = [3L/10, 2W/5, 0]^T$ and  ${\bm \psi}_{\rm t}^{\rm III} = [-3L/10, 3W/5, 0]^T$, respectively. Meanwhile, to validate the effectiveness of the proposed algorithm, the exhaustive search-based algorithm is adopted as a baseline, which searches the location of each TPA in the area $[-L/2, L/2]$ with the step size of $0.5~{\rm m}$. 
	
	\begin{figure}[htbp]
		\centering
		\subfloat[Communication rate obtained by the proposed algorithm and the exhaustive search-based algorithm versus the radar SNR requirement.]{\includegraphics[width=0.95\columnwidth]{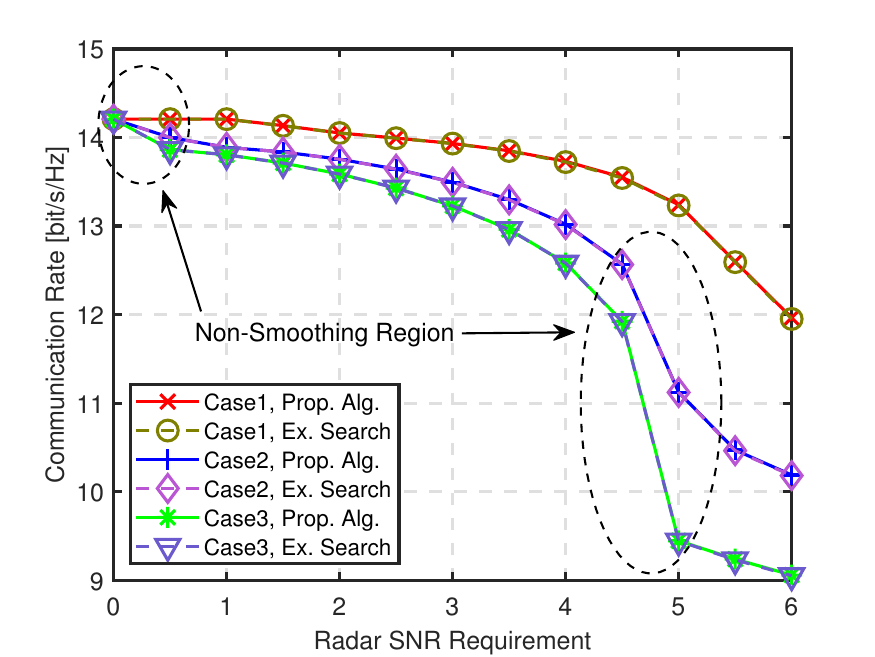}}
		\hfill
		\subfloat[Optimized horizontal coordinates obtained by the proposed algorithm of TPAs versus the radar SNR requirement under each case. ]{\includegraphics[width=0.95\columnwidth]{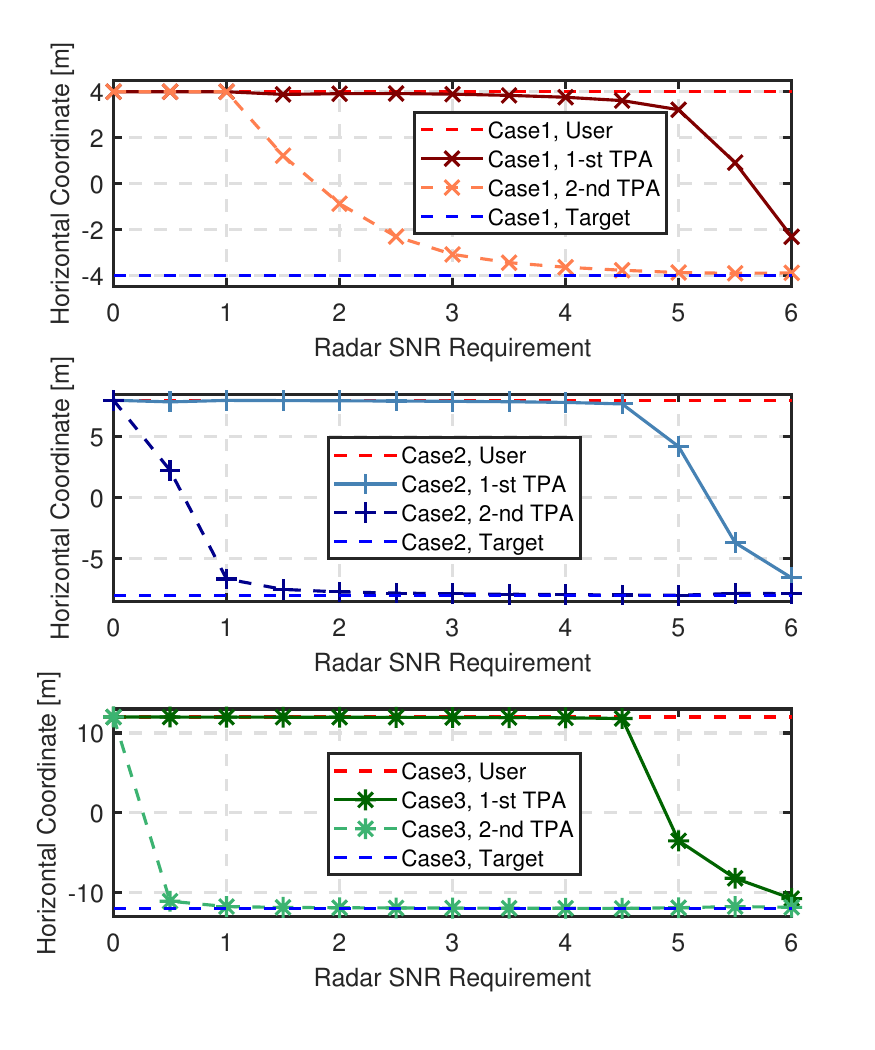}}
		\hfill
		\caption{Comparison between the proposed algorithm and the exhaustive search-based algorithm with $M =2$ under Case1, Case2, and Case3.  }
		\label{fig: special_case}
	\end{figure}

    
	Fig. \ref{fig: special_case}(a) compares the proposed algorithm (marked by Prop. Alg.) and the exhaustive search-based algorithm (marked by Ex. Search) in terms of the communication rate versus the radar SNR requirement under Case1, Case2, and Case3. For each case, one can observe that the results obtained by the proposed algorithm and the exhaustive search-based algorithm are very close, which validates the effectiveness of the proposed algorithm. Besides, the communication rate under Case1 tends to decrease gradually and smoothly with the increment of radar SNR requirement. But under Case2 and Case3,  two \emph{non-smoothing regions} (characterized by abrupt rate drops) are observed. The regions correspond to the critical communication-sensing trade-off in pinching-antenna ISAC systems. 
    
    Analyzing the underlying causes of \emph{non-smoothing regions} is  fundamental to unraveling the operational mechanisms of pinching antennas. To this end, Fig. \ref{fig: special_case}(b) illustrates the optimal horizontal coordinates of TPAs (i.e., $x_{{\rm T}, 1}^{\star}$ and $x_{{\rm T}, 2}^{\star}$) versus the radar SNR requirement under the three cases. For each case, the increment of radar SNR requirement requires the two TPAs to approach the target sequentially to reduce the path loss for sensing.  More specifically, for Case2 and Case3, it is observed that the $2$-nd TPA intends to depart from the user when the radar SNR requirement increases from $0$ to $0.5$, while the $1$-st TPA intends to depart from the user when the radar SNR requirement increases from $4.5$ to $5$. This behavior aligns with the two \emph{non-smoothing regions} in Fig. \ref{fig: special_case}(a). The departure of TPAs from the user greatly enlarges TPA-user path loss (inverse-square with respect to the transmission distance), and thus, induces a  significant degradation in the communication rate. Furthermore, comparing the three cases reveals that a larger user-target distance not only complicates the communication-sensing trade-off but also causes the emergence of the \emph{non-smoothing regions}. 
	
	\subsection{Evaluation of Pinching-Antenna ISAC Systems}
	
	In this subsection, we evaluate the performance of pinching-antenna  ISAC systems via the comparison with the following benchmarks:
	\begin{itemize}
		\item Conventional fixed-antenna system: This setting implements each TPA at the center of the serving area, i.e., $x_{{\rm T}, m} = 0$, $\forall m \in \mathcal{M}$.
		\item User-centric antenna system: This setting aims to enhance the communication performance by deploying the TPAs near the user, i.e., $x_{{\rm T}, m} = x_{\rm u}$, $\forall m \in \mathcal{M}$.
		\item Target-oriented antenna system: This setting aims to improve the probing range by deploying the TPAs near the target, i.e., $x_{{\rm T}, m} = x_{\rm t}$, $\forall m \in \mathcal{M}$.
		\item Midpoint antenna system: This setting aims to balance the communication-sensing trade-off  by deploying the TPAs at the midpoint between the user and the target, i.e., $x_{{\rm T}, m} = (x_{\rm u}+x_{\rm t})/2$, $\forall m \in \mathcal{M}$.
	\end{itemize}
	
	To mitigate the impact of randomness caused by the locations of user and target on the system performance, multiple Monte Carlo realizations have been conducted with results averaged over all realizations. It is noted that if Problem ${\bf P}_1$ is infeasible to one realization, the corresponding communication rate is marked as $0$.
	
	\begin{figure}
		\begin{center}
			\centerline{\includegraphics[ width=.49\textwidth]{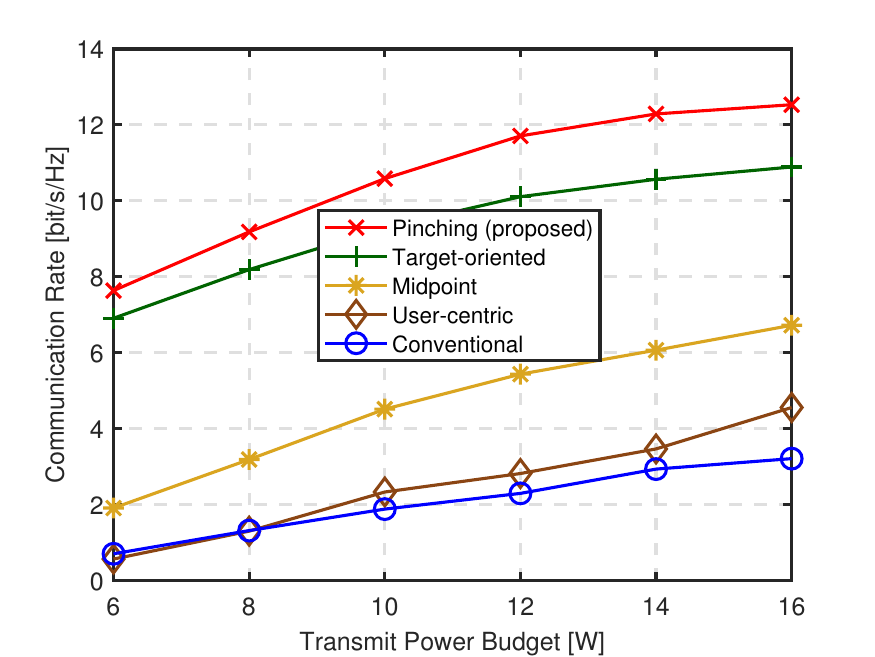}}
			\caption{Comparison among the proposed pinching, the target-oriented, the midpoint, the user-centric, and the conventional antenna systems.}
			\label{fig: p_max}
		\end{center}
		\vspace{-0.4cm}
	\end{figure}
	
	Fig. \ref{fig: p_max} compares the pinching antenna system with the benchmarks in terms of communication performance versus the transmit power budget. It is observed that the pinching antenna system achieves the highest communication rate among all systems, and the performance gain goes larger with the transmit power budget. The reason is that the pinching antenna system facilitates the flexible large-scale adjustment of TPAs according to the locations of  the user and the target, which not only reduces the propagation loss but also provides a new degree of freedom to mitigate the communication-sensing interference. Moreover, the joint optimization of deployment and beamforming design of pinching antennas inherently outperforms the benchmarks that only optimize beamforming.  

	
	\begin{figure}
		\begin{center}
			\centerline{\includegraphics[ width=.49\textwidth]{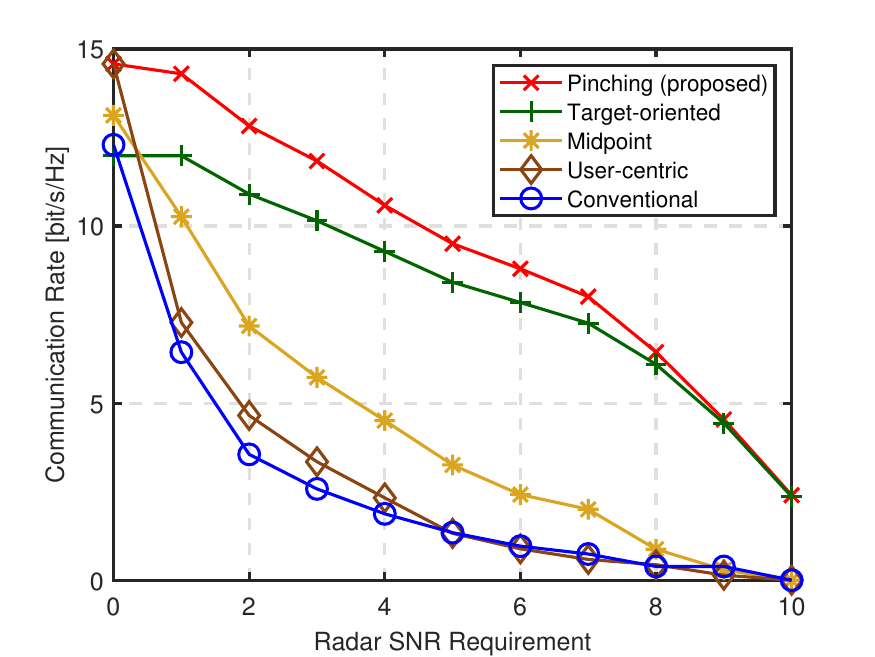}}
			\caption{Comparison among the proposed pinching, the target-oriented, the midpoint, the user-centric, and the conventional antenna systems.}
			\label{fig: snr}
		\end{center}
		\vspace{-0.4cm}
	\end{figure}
	
	Fig. \ref{fig: snr} compares the  pinching antenna system with the benchmarks in terms of communication-sense region characterized by communication rate and radar SNR. One can find that the  pinching antenna system obtains the largest region that encompasses the regions achieved by benchmarks, attributed to  the extra spatial degree of freedom brought by movable TPAs. Besides, as observed, the target-oriented design achieves similar performance to the pinching antenna design for high radar SNR requirement, since the TPAs should be deployed near the target to guarantee it within the probing range. The user-centric design is competitive to the pinching antenna design for low radar SNR requirement, since it minimizes the TPA-user distance. Moreover, there exists a significant gap between the midpoint  design and the pinching antenna design, which necessitates the optimization of TPA locations.

    
	\begin{figure}
		\begin{center}
			\centerline{\includegraphics[ width=.49\textwidth]{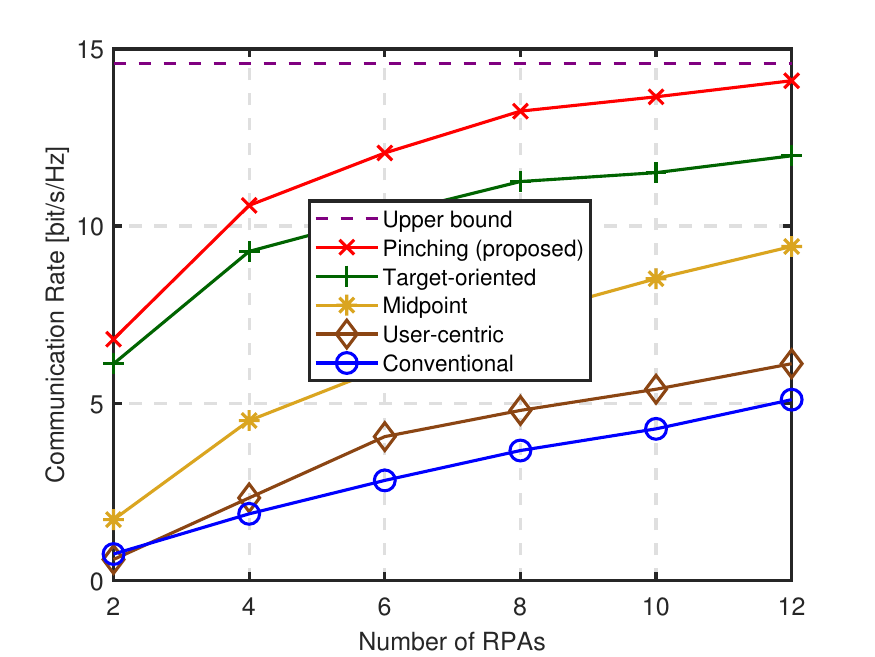}}
			\caption{Impact of the number of RPAs on the pinching-antenna ISAC systems.}
			\label{fig: n}
		\end{center}
		\vspace{-0.4cm}
	\end{figure}
	
	Fig. \ref{fig: n} shows the impact of the number of RPAs on the system performance. It is observed that the communication rate obtained by all the systems increases with the number of RPAs, as more RPAs can enhance the receiving antenna gain and reduce the average RPA-target path loss. Nevertheless, the system performance cannot be infinitely improved by increasing the number of RPAs, and there exists a upper bound  (about $14.57~{\rm bit/s/Hz}$ in this case), which corresponds to solving Problem ${\bf P}_1$ with $\Gamma_{\rm Req} = 0$. Besides, the pinching antenna system is superior to benchmarks.

	\begin{figure}
		\begin{center}
			\centerline{\includegraphics[ width=.49\textwidth]{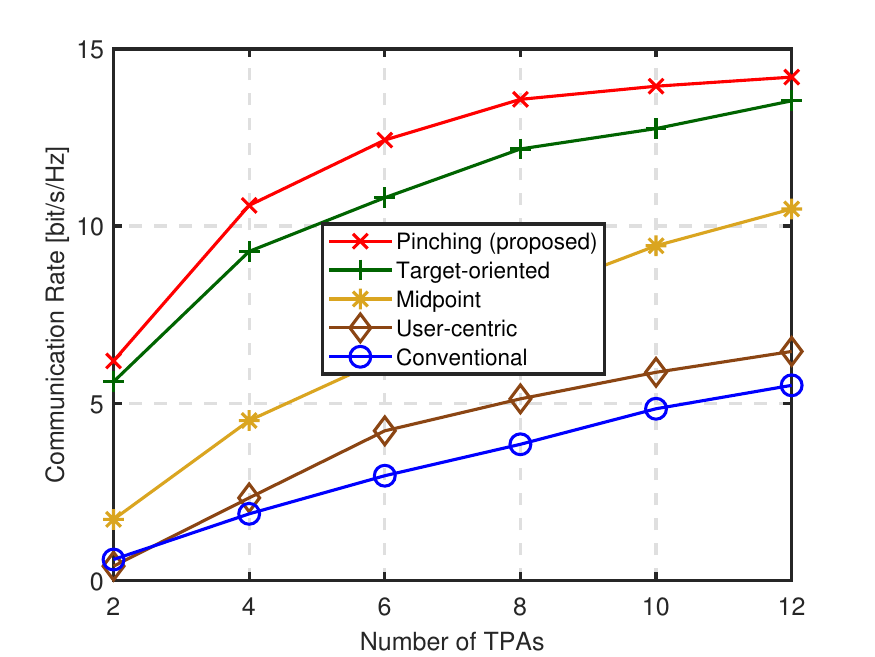}}
			\caption{Impact of the number of TPAs on pinching-antenna ISAC systems.}
			\label{fig: m}
		\end{center}
		\vspace{-0.4cm}
	\end{figure}
	
	Fig. \ref{fig: m} shows the impact of the number of TPAs on the system performance. One can find that the communication performance can be improved by increasing the number of TPAs, and the performance gain is slightly greater than that obtained by increasing the number of RPAs (cf. Fig. \ref{fig: n}). The reason is that more TPAs can not only enhance the transmit antenna gain and reduce the average propagation distance to both user and target, but also bring more spatial degrees of freedom for joint deployment and  beamforming design of TPAs, resulting in a better performance gain. Furthermore, the system performance exhibits diminishing marginal gains with the increasing number of TPAs or RPAs, suggesting that when partitioning a pinching antenna array, the numbers of TPAs and RPAs should be balanced.

	\section{Conclusion}
    This paper has investigated a multi-waveguide pinching-antenna ISAC system with the aim to simultaneously detect one potential target and  serve one downlink information user. We formulated a communication rate maximization problem under the constraints of radar SNR requirement, transmit power budget, and movable area of TPAs. A novel fine-tuning method was proposed to approximate the non-convex problem with negligible performance loss. Then, the approximated problem was solved by an SCA-based algorithm. Numerical results evaluated the proposed SCA-based algorithm in comparison with the exhaustive search-based algorithm. Besides, pinching-antenna ISAC systems were shown to be superior to conventional systems and three pinching-antenna benchmarks.     Our study on multi-waveguide pinching-antenna ISAC systems yielded  insightful observations, including {\emph{non-smoothing regions}} in communication rate as the radar SNR requirement increases, and diminishing marginal performance gains with the number of TPAs or RPAs. 


    \appendix
    
    \subsection{Proof of Proposition 3}
    
    We first prove that the influence on distance caused by fine tuning is no more than $\lambda$.  
		
		It is observed that $x_{{\rm T}, m}^{\star}$ should be between $x_{\rm u}$ and ${x}_{\rm t}$. Otherwise, by moving $x_{{\rm T}, m}^{\star}$ to $x_{\rm u}$ (or ${x}_{\rm t}$), both communication and sensing performance can be enhanced.
		
		Denote $D_{{\rm t},m}^{\star} \triangleq || {\bm \psi}_{\rm t} -  {\bm \psi}_{{\rm T}, m}^{\star} ||$ and $D_{{\rm u},m}^{\star} \triangleq || {\bm \psi}_{\rm u} -  {\bm \psi}_{{\rm T}, m}^{\star} ||$ by the distance before fine tuning, $\widehat{D}_{{\rm t},m}^{\star} \triangleq || {\bm \psi}_{\rm t} -  \widehat{\bm \psi}_{{\rm T}, m}^{\star} ||$ and $\widehat{D}_{{\rm u},m}^{\star} \triangleq || {\bm \psi}_{\rm u} -  \widehat{{\bm \psi}}_{{\rm T}, m}^{\star} ||$ by the distance after fine tuning. Due to the continuity of $\theta_m ( x_{{\rm T}, m} ) $ with respect to $x_{{\rm T}, m}$, there exists $\widehat{x}_{{\rm T}, m}^{\star}$ such that $|\theta_m \left( \widehat{x}_{{\rm T}, m}^{\star} \right) - \theta_m \left( {x}_{{\rm T}, m}^{\star} \right) | \leq 1$
		and \eqref{re_theta} holds, which indicates that
		\begin{flalign}
			\lambda & \geq \left| \widehat{D}_{{\rm t},m}^{\star} - \widehat{D}_{{\rm u},m}^{\star} -\left( {D}_{{\rm t},m}^{\star} - {D}_{{\rm u},m}^{\star}     \right)           \right| \nonumber \\
			& \geq {\max} \left\{ \left|\widehat{D}_{{\rm t},m}^{\star} - {D}_{{\rm t},m}^{\star} \right|, \left|\widehat{D}_{{\rm u},m}^{\star} - {D}_{{\rm u},m}^{\star} \right|          \right\}. \label{appro_error}
		\end{flalign}
		Assume that $\|{\bf h}_{\rm t}^{\star}\|=\|\widehat{\bf h}_{\rm t}^{\star}\|$ and $\|{\bf h}_{\rm u}^{\star}\|=\|\widehat{\bf h}_{\rm t}^{\star}\|$, we prove that $\{\widehat{c}^{\star}_{\rm u}, \widehat{c}^{\star}_{\rm t}, \widehat{x}_{{\rm T}, m}^{\star}   \}$ satisfies the constraints \eqref{p1d}, \eqref{p2b}, and \eqref{p2c}. With $\{\widehat{c}^{\star}_{\rm u}, \widehat{c}^{\star}_{\rm t}, \widehat{x}_{{\rm T}, m}^{\star}   \}$, we can obtain an objective value which is not inferior to $\{{c}^{\star}_{\rm u}, {c}^{\star}_{\rm t}, {x}_{{\rm T}, m}^{\star}   \}$.
		
		As shown in \eqref{appro_error}, the movement of $\{ \widehat{x}_{{\rm T},m}^{\star} \}$ is at a scale of wavelength, which can be ignored when compared with $L$. Therefore, the constraint \eqref{p1d} holds.
		
		Next, we show the existence of $\zeta$ (i.e., the solution to \eqref{zeta}) by proving that
		\begin{flalign}
			& \left| \left({\bf h}_{\rm t}^{\star} \right)^H {\bf h}_{\rm u}^{\star} \right| = \left| \sum_{m \in \mathcal{M}} \frac{\eta e^{ 2 \pi j \theta_m \left( x_{{\rm T}, m}^{\star} \right)}}{ D_{{\rm u}, m}^{\star}  D_{{\rm t}, m}^{\star} }  \right| \leq \sum_{m \in \mathcal{M}}   \frac{\eta }{ D_{{\rm u}, m}^{\star}  D_{{\rm t}, m}^{\star} }  \nonumber \\
			& =  \left| \sum_{m \in \mathcal{M}} \frac{\eta e^{ 2 \pi j \theta_m \left( \widehat{x}_{{\rm T}, m}^{\star} \right)}}{ \widehat{D}_{{\rm u}, m}^{\star}  \widehat{D}_{{\rm t}, m}^{\star} }  \right| = \left| \left(\widehat{\bf h}_{\rm t}^{\star} \right)^H \widehat{\bf h}_{\rm u}^{\star} \right|. \label{tri}
		\end{flalign}
		Thus, we can find $\zeta$ such that
		\begin{flalign}
			{\rm Re} \left\{ c_{\rm u}^{\star}  \left(c_{\rm t}^{\star}\right)^{*}  \left({\bf h}_{\rm t}^{\star} \right)^{H} {\bf h}_{\rm u}^{\star}  \right\}  = {\rm Re} \left\{   \widehat{c}_{\rm u}^{\star}  \left( \widehat{c}_{\rm t}^{\star}  \right)^{*} \left(\widehat{\bf h}_{\rm t}^{\star} \right)^H \widehat{\bf h}_{\rm u}^{\star}  \right\}           \label{re_equal},
		\end{flalign}
        which is equivalent to equation \eqref{zeta}.

		Combining $|\widehat{c}_{\rm u}^{\star} | = |{c}_{\rm u}^{\star}  |$, $|\widehat{c}_{\rm t}^{\star} | = |{c}_{\rm t}^{\star}  |$, and \eqref{re_equal}, one can find 
		\begin{flalign}
			f_{\rm p}\left( \widehat{c}_{\rm u}^{\star}, \widehat{c}_{\rm t}^{\star}, \{ \widehat{x}_{{\rm T}, m}^{\star}\}   \right) = f_{\rm p}\left( c_{\rm u}^{\star}, c_{\rm t}^{\star}, \{x_{{\rm T}, m}^{\star}\}   \right) \leq P_{\max},
		\end{flalign}
		which indicates that \eqref{p2b} holds. 
		
		Besides, by substituting \eqref{tri} and \eqref{re_equal} into the functions	$f_{\rm t}\left( \widehat{c}_{\rm u}^{\star}, \widehat{c}_{\rm t}^{\star}, \{\widehat{x}_{{\rm T}, m}^{\star} \}  \right)$ and $f_{\rm u}\left( \widehat{c}_{\rm u}^{\star}, \widehat{c}_{\rm t}^{\star}, \{\widehat{x}_{{\rm T}, m}^{\star}\}   \right)$, we reach
		\begin{flalign}
			&f_{\rm t}\left( \widehat{c}_{\rm u}^{\star}, \widehat{c}_{\rm t}^{\star}, \{ \widehat{x}_{{\rm T}, m}^{\star} \}  \right) \geq f_{\rm t}\left( c_{\rm u}^{\star}, c_{\rm t}^{\star}, \{ x_{{\rm T}, m}^{\star}\}   \right) \geq \frac{\Gamma_{\rm Req} \sigma_{\rm s}^2}{\beta}, \\
			&f_{\rm u}\left( \widehat{c}_{\rm u}^{\star}, \widehat{c}_{\rm t}^{\star}, \{ \widehat{x}_{{\rm T}, m}^{\star} \}  \right) \geq f_{\rm u}\left( c_{\rm u}^{\star}, c_{\rm t}^{\star}, \{ x_{{\rm T}, m}^{\star} \}  \right) ,
		\end{flalign}
		which shows that $\{\widehat{c}^{\star}_{\rm u}, \widehat{c}^{\star}_{\rm t}, \widehat{x}_{{\rm T}, m}^{\star}   \}$ satisfies the constraint \eqref{p2c} and obtains a objective value which is not inferior to that obtained by $\{c_{\rm u}^{\star}, c_{\rm t}^{\star}, x_{{\rm T}, m}^{\star} \}$.

\end{document}